\documentclass[
    aps,
    prx,
    floatfix,
    reprint,
    superscriptaddress,
]{revtex4-2}

\usepackage{graphicx}
\usepackage{amssymb}
\usepackage{amsmath}
\usepackage{braket}

\usepackage{bibunits}

\usepackage{xcolor}
\definecolor{linkblue}{RGB}{20, 55, 90}
\usepackage[
    colorlinks=true,
    linkcolor=linkblue,
    citecolor=blue,
    urlcolor=blue,
    bookmarks=true,
    bookmarksopen=true,
    bookmarksnumbered=true,
]{hyperref}
\usepackage{bookmark}

\defaultbibliographystyle{apsrev4-2-etal}
\defaultbibliography{references.bib}

\newcommand{\mainnumbering}{
    \setcounter{figure}{0}
    \setcounter{table}{0}
    \setcounter{equation}{0}

    \renewcommand{\thefigure}{\arabic{figure}}
    \renewcommand{\thetable}{\arabic{table}}
    \renewcommand{\theequation}{\arabic{equation}}

    \renewcommand{\theHfigure}{main.\arabic{figure}}
    \renewcommand{\theHtable}{main.\arabic{table}}
    \renewcommand{\theHequation}{main.\arabic{equation}}
}

\newcommand{\methodsnumbering}{
    \setcounter{figure}{0}
    \setcounter{table}{0}
    \setcounter{equation}{0}

    \renewcommand{\thefigure}{M\arabic{figure}}
    \renewcommand{\thetable}{M\arabic{table}}
    \renewcommand{\theequation}{M\arabic{equation}}

    \renewcommand{\theHfigure}{methods.\arabic{figure}}
    \renewcommand{\theHtable}{methods.\arabic{table}}
    \renewcommand{\theHequation}{methods.\arabic{equation}}
}

\newcommand{\suppnumbering}{
    \setcounter{section}{0}
    \setcounter{subsection}{0}
    \setcounter{figure}{0}
    \setcounter{table}{0}
    \setcounter{equation}{0}

    \renewcommand{\thesection}{S\arabic{section}}
    \renewcommand{\thesubsection}
        {\thesection.\arabic{subsection}}

    \renewcommand{\thefigure}{S\arabic{figure}}
    \renewcommand{\thetable}{S\arabic{table}}
    \renewcommand{\theequation}{S\arabic{equation}}

    \renewcommand{\theHsection}{supp.\arabic{section}}
    \renewcommand{\theHsubsection}
        {supp.\arabic{section}.\arabic{subsection}}
    \renewcommand{\theHfigure}{supp.\arabic{figure}}
    \renewcommand{\theHtable}{supp.\arabic{table}}
    \renewcommand{\theHequation}{supp.\arabic{equation}}
}

\makeatletter

\newcommand{\methodsection}[1]{
    \bookmarksetupnext{level=subsection}
    \section{#1}
}

\newcommand{\methodsubsection}[1]{
    \bookmarksetupnext{level=subsubsection}
    \subsection{#1}
}

\newcommand{\supplementcontents}{
    \bookmarksetupnext{level=subsection}
    \section*{Contents}
    \@starttoc{stoc}
}

\newcommand{\suppsection}[1]{
    \bookmarksetupnext{level=subsection}
    \section{#1}

    \addcontentsline{stoc}{section}{
        \protect\numberline{\thesection}#1
    }
}

\newcommand{\suppsubsection}[1]{
    \bookmarksetupnext{level=subsubsection}
    \subsection{#1}

    \addcontentsline{stoc}{subsection}{
        \protect\numberline{\thesubsection}#1
    }%
}

\makeatother

\newcommand{\majorsection}[2]{
    \par
    \hypertarget{#2}{}
    \bookmark[level=section,dest=#2]{#1}
    \vspace{1em}
    \begin{center}
        {\Large\bfseries #1}
    \end{center}
    \vspace{0em}
}

\makeatletter

\newcommand{\setbiblinksuffix}[1]{%
    \gdef\@extra@binfo{#1}%
    \gdef\@extra@b@citeb{#1}%
}

\makeatother

\usepackage{bm}

\def \B{{\bm B}}
\def \J{{\bm J}}
\def \R{{\bm R}}

\def \g{{\bm g}}

\def \q{{\bm q}}
\def \u{{\bm u}}
\def \x{{\bm x}}

\def \z{{\bm z}}

\let\txtr\r
\let\txtk\k

\def\r{\ifmmode{\bm r}\else\expandafter\txtr\fi}
\def\k{\ifmmode{\bm k}\else\expandafter\txtk\fi}

\begin{document}

\title{
\texorpdfstring{
    Quantum noise spectroscopy of superconducting dynamics in a cuprate thin film
}{
    Main
}
}

\author{Zhongyuan~Liu}
\thanks{These authors contributed equally to this work.}
\affiliation{Department of Physics, Washington University, St. Louis, Missouri 63130, USA}

\author{Ruotian~Gong}
\thanks{These authors contributed equally to this work.}
\affiliation{Department of Physics, Washington University, St. Louis, Missouri 63130, USA}

\author{Jaewon~Kim}
\thanks{These authors contributed equally to this work.}
\affiliation{Department of Physics, University of California, Berkeley, California 94720, USA}

\author{Oriana~K.~Diessel}
\thanks{These authors contributed equally to this work.}
\affiliation{ITAMP, Center for Astrophysics, Harvard \& Smithsonian, Cambridge, Massachusetts 02138, USA}
\affiliation{Department of Physics, Harvard University, Cambridge, Massachusetts 02138, USA}

\author{Jiajie~Cheng}
\thanks{These authors contributed equally to this work.}
\affiliation{Department of Physics, Carnegie Mellon University, Pittsburgh, Pennsylvania 15213, USA}

\author{Qiaozhi~Xu}
\affiliation{Department of Physics, Washington University, St. Louis, Missouri 63130, USA}

\author{Zackary~Rehfuss}
\affiliation{Department of Physics, Washington University, St. Louis, Missouri 63130, USA}

\author{Xinyi~Du}
\affiliation{Department of Physics, Washington University, St. Louis, Missouri 63130, USA}

\author{Guanghui~He}
\affiliation{Department of Physics, Washington University, St. Louis, Missouri 63130, USA}

\author{Abhishek~Singh}
\affiliation{Department of Physics and Astronomy, Texas Tech University, Lubbock, Texas 79409, USA}

\author{Yun~Suk~Eo}
\affiliation{Department of Physics and Astronomy, Texas Tech University, Lubbock, Texas 79409, USA}

\author{Erik~A.~Henriksen}
\affiliation{Department of Physics, Washington University, St. Louis, Missouri 63130, USA}
\affiliation{Institute of Materials Science and Engineering, Washington University, St. Louis, Missouri 63130, USA}

\author{G.~D.~Gu}
\affiliation{Condensed Matter Physics and Materials Science Department, Brookhaven National Laboratory, Upton, New York 11973, USA}

\author{Norman~Y.~Yao}
\affiliation{Department of Physics, Harvard University, Cambridge, Massachusetts 02138, USA}

\author{Francisco~Machado}
\affiliation{ITAMP, Center for Astrophysics, Harvard \& Smithsonian, Cambridge, Massachusetts 02138, USA}
\affiliation{Department of Physics, Harvard University, Cambridge, Massachusetts 02138, USA}

\author{Sheng~Ran}
\affiliation{Department of Physics, Washington University, St. Louis, Missouri 63130, USA}
\affiliation{Institute of Materials Science and Engineering, Washington University, St. Louis, Missouri 63130, USA}

\author{Shubhayu~Chatterjee}
\email{shubhayuchatterjee@cmu.edu}
\affiliation{Department of Physics, Carnegie Mellon University, Pittsburgh, Pennsylvania 15213, USA}

\author{Chong~Zu}
\email{zu@wustl.edu}
\affiliation{Department of Physics, Washington University, St. Louis, Missouri 63130, USA}
\affiliation{Institute of Materials Science and Engineering, Washington University, St. Louis, Missouri 63130, USA}

\begin{abstract}
Characterizing the low-energy dynamics of quantum materials is crucial for understanding strongly correlated electronic states, 
yet remains experimentally challenging, particularly in two-dimensional systems.
Here, we leverage nitrogen-vacancy (NV) centers in diamond as a non-invasive quantum sensor to study thin-film Bi$_2$Sr$_2$CaCu$_2$O$_{8+\delta}$ (BSCCO), revealing several distinct dynamical phenomena across the superconducting phase diagram.
At zero magnetic field and low temperatures, NV depolarization ($T_1$) spectroscopy captures GHz-scale noise from nodal quasiparticle excitations, consistent with Bardeen--Cooper--Schrieffer (BCS) mean-field theory.
Near the critical temperature $T_c \approx 90$~K, supercurrent-fluctuation-induced noise sharply reduces $T_1$.
By analyzing the temperature scaling of $T_1$, we observe clear deviations from mean-field prediction, reflecting the importance of order parameter fluctuations and enabling the determination of critical exponents.
Under a small magnetic field, a broad and asymmetric reduction of $T_1$ unveils a liquid-like phase of mobile vortices.
NV decoherence ($T_2$) spectroscopy further probes MHz-scale noise and reveals complex vortex-solid fluctuations well below $T_c$.
Our results establish quantum noise spectroscopy as a non-invasive probe of superconducting dynamics over frequency and length scales complementary to existing techniques.
\end{abstract}

\date{\today}

\maketitle


\mainnumbering

\setbiblinksuffix{-main}
\begin{bibunit}[apsrev4-2-etal]

\emph{Introduction ---}
Strong electronic correlations are responsible for the plethora of complex emergent phenomena observed in quantum materials.
This is perhaps best exemplified in unconventional superconductors, where electronic interactions induce a macroscopic quantum coherence of electron pairs and dissipationless flow of current \cite{tinkham2004introduction,keimer2015quantum,zhou2021high}.
Key features of this state are encoded in the low-energy dynamics, manifesting as quasiparticle excitations arising from the breaking of electron-pairs \cite{bardeen1957microscopic,bardeen1957theory,bogoljubov1958new}, slow collective modes such as critical pairing fluctuations near the metal-superconductor phase transition \cite{larkin2005theory,schmid1966time,maki1968critical,cyrot1973ginzburg,schuller2006time}, or topological excitations such as vortices carrying quantized magnetic flux \cite{tinkham2004introduction,blatter1994vortices,halperin1979resistive,beasley1979possibility,nelson1988vortex}.
Characterizing these dynamics thus offers a new avenue towards understanding the superconducting state.
This is particularly important when addressing the emerging landscape of correlated, two-dimensional superconducting materials \cite{balents2020superconductivity,ko2025signatures}, where the application of conventional experimental probes is a notorious challenge.

\begin{figure*}[t]
\centering
\includegraphics[width=\textwidth]{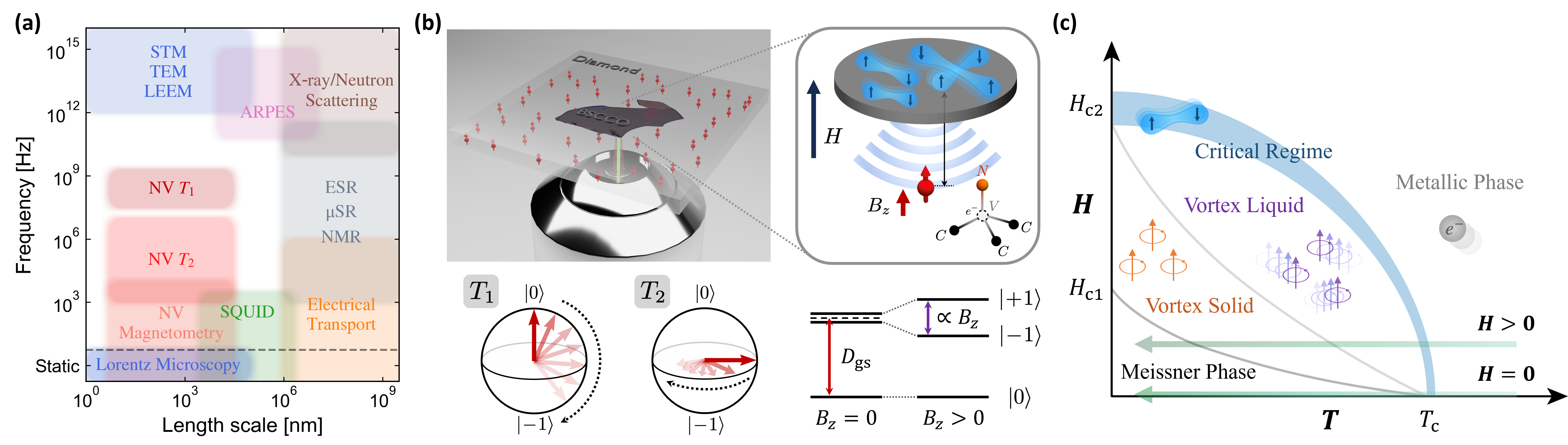}
\caption{{\bf NV noise spectroscopy of superconducting dynamics.}
(a) Summary of different experimental techniques for probing superconductivity, organized by their frequency and length scales, highlighting the complementarity of our qubit-based noise spectroscopy (NV $T_1$ and $T_2$) and DC magnetometry~\cite{barry2020sensitivity}.
For comparison, these techniques are mapped to equivalent frequencies corresponding to the characteristic energy scales of the probed phenomena.
Techniques depicted include: 
electrical transport~\cite{essaleh2018theoretical}, 
superconducting quantum interference device (SQUID)~\cite{kirtley1999scanning,dusad2019magnetic},
Lorentz microscopy~\cite{harada1992realtime,silhanek2022local}, 
X-ray/neutron scattering~\cite{comin2016resonant,muhlbauer2019magnetic}, 
magnetic resonance with nuclear, electronic, or muonic spins (NMR/ESR/µSR)~\cite{hore2015nuclear,hillier2022muon}, 
angle-resolved photoemission spectroscopy (ARPES)~\cite{liu2008development, kitamura2022development}, 
low-energy and transmission electron microscopy (LEEM/TEM)~\cite{altman2010trends,moradifar2023accelerating}, 
and scanning tunneling microscope (STM)~\cite{parker2010nanoscale}.
Some techniques, while capable of accessing high momentum data, require macroscopic sample volumes and, thus, are only suitable for probing mm-scale samples.
Techniques like pump-probe spectroscopy \cite{kaindl2000ultrafast,delatorre2021colloquium}, which are known to drive the sample strongly out of equilibrium, are not included.
(b) Schematic of the experimental setup: a thin film ($\sim 200$~nm thickness) of BSCCO is directly exfoliated onto a diamond plate containing a thin layer of NV ensemble beneath its surface.
Current fluctuations in the BSCCO sample induce magnetic field noise at the NV locations, which can be detected by either depolarization ($T_1$) or decoherence ($T_2$) noise spectroscopy.
An out-of-plane magnetic field $H$ is applied, which is parallel to the quantization axis of the [111] NV group.
The NV exhibits a splitting of its $|m_s=\pm1\rangle$ sublevels, which is determined by the local magnetic field $B_z$, centered around a zero-field splitting $D_\mathrm{gs} = 2.87$~GHz.
(c) Schematic $H$--$T$ phase diagram of a typical type-II superconductor \cite{tinkham2004introduction,keimer2015quantum,zhou2021high}.
The green arrows indicate zero-field cooling ($H = 0$) and field-cooling ($H > 0$) pathways used in our experiments. 
}
\label{fig:fig1}
\end{figure*}

Over the past decades, several experimental techniques have been developed to probe correlated quantum dynamics in superconductors~\cite{comin2016resonant,kaindl2000ultrafast,delatorre2021colloquium,parker2010nanoscale,essaleh2018theoretical,hore2015nuclear}.
Broadly, such techniques may be categorized into scattering methods, local spectroscopic probes, and bulk measurements [Fig.~\ref{fig:fig1}(a)].
Scattering probes (e.g., X-ray scattering~\cite{comin2016resonant}) provide both energy- and momentum-resolved data, but they typically require experimental apparatus beyond table-top settings, large sample volumes, and can drive the system out of equilibrium~\cite{kaindl2000ultrafast,delatorre2021colloquium}.
Local probes such as scanning tunneling microscopy (STM)~\cite{parker2010nanoscale} are applicable to low-dimensional samples and provide lattice-scale spatial resolution, yet they exhibit limited momentum resolution and typically operate at low temperatures.
By contrast, bulk measurements (including transport~\cite{essaleh2018theoretical} and nuclear magnetic resonance~\cite{hore2015nuclear}) can operate over a wide temperature range, but generally demand a large sample volume.
Therefore, resolving spectral features at long wavelengths below the meV scale in two-dimensional materials and thin film samples remains a challenge, leading to an unfulfilled gap for experimental techniques that can probe equilibrium dynamics at both low energies and a variety of length scales.

In this work, we fill this gap by employing a non-invasive, table-top experimental technique --- quantum noise spectroscopy --- to probe a wide range of low-energy dynamical phenomena in superconductors.
Quantum noise spectroscopy has already been successfully deployed to probe electronic charge and spin dynamics in myriad platforms, including metals~\cite{kolkowitz2015probing, hsieh2019imaging}, semi-metals~\cite{andersen2019electron}, and insulating van der Waals magnets~\cite{mclaughlin2022quantum, ziffer2024quantum, melendez2025quantum}.
However, its application in the study of superconductivity has been more limited; previous studies have primarily utilized qubit spins to image static magnetic signatures, such as the Meissner effect and vortices induced by applied fields~\cite{bouchard2011detection, schlussel2018wide, xu2019mapping, bhattacharyya2024imaging}.
Here, we instead utilize the spin relaxation time ($T_1$) and coherence time ($T_2$) to directly capture the spectral density of magnetic noise generated by dynamical processes in superconductors, rather than imaging their static properties.
Using NV centers positioned in close proximity to a thin film high-$T_c$ cuprate, Bi$_2$Sr$_2$CaCu$_2$O$_{8+\delta}$ (BSCCO, Bi-2212), we characterize superconducting fluctuations as a function of temperature, frequency, and applied magnetic field.
We demonstrate in-situ detection of spatially resolved dynamics spanning a broad frequency range (MHz to GHz) and length scales (nm to $\mathrm{\mu m}$), thus offering a powerful complementary toolset to existing techniques for probing superconductivity [Fig.~\ref{fig:fig1}(a)].

Our results are threefold.
First, at zero applied field, we observe and characterize the fluctuations arising from both long-lived quasiparticle excitations at low temperature and collective critical modes near the metal-superconductor transition temperature $T_c$. 
These fluctuations induce GHz-scale magnetic noise at the location of the nearby NVs, which can be characterized via NV depolarization ($T_1$) spectroscopy.
These capabilities enable us to directly measure the characteristic timescale $\tau$ of critical slowing down of pairing fluctuations near $T_c$ and determine both the static and dynamical critical exponents associated with the phase transition \cite{landau1980statistical,lifshitz1980statistical,hohenberg1977theory}. 

Second, we directly measure the broadening of the transition in the presence of a small applied field $H$.
At non-zero $H$, the system also hosts superconducting vortices whose dynamics induce additional magnetic field fluctuations.
The field dependence of the NV relaxation rate is consistent with the existence of an intermediate-temperature vortex liquid phase \cite{blatter1994vortices,halperin1979resistive} where magnetic flux lines diffuse through the superconducting film, and further allows us to extract an effective vortex-pinning scale.

Finally, we employ decoherence ($T_2$) noise spectroscopy to study lower-frequency fluctuations in the MHz regime~\cite{machado2023quantum}.
Deep within the superconducting phase, we observe a distinct noise mechanism that significantly enhances the NV decoherence rate. 
These strong, low-frequency magnetic fluctuations potentially arise from thermally assisted flux motion in a vortex solid phase~\cite{blatter1994vortices,fisher1989vortex,fisher1991thermal}.

\begin{figure*}[t]
    \centering
    \includegraphics[width=0.85\textwidth]{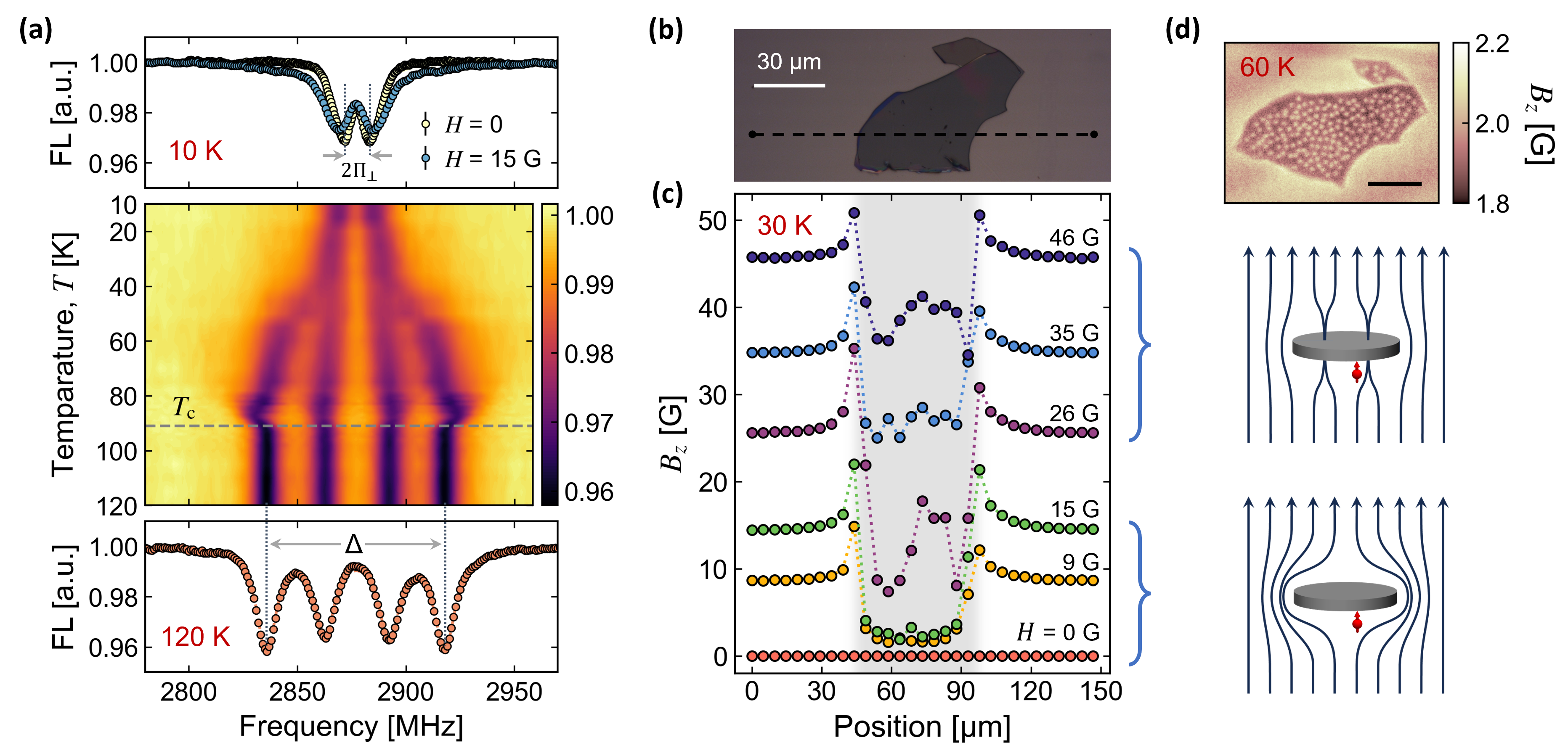}
    \caption{{\bf Observing the Meissner effect and superconducting vortices.}
    (a) ODMR spectrum of NV centers under BSCCO with an applied field $H = 15$~G at temperatures from 10~K to 120~K.
    The colors represent the normalized fluorescence contrast (FL) of the spectrum.
    The gray dashed line marks the superconducting transition temperature $T_c \approx 90$~K.
    Top: at 10~K, ODMR spectrum of NV centers under BSCCO at $H=0$ (yellow), where the intrinsic splitting $2\Pi_\perp$ originates from the local electric field~\cite{mittiga2018imaging}.
    At $H = 15$~G, the spectrum (blue) remains largely unchanged, indicating a strong magnetic field repulsion due to the Meissner effect.
    Bottom: when the bias field is not suppressed (120~K), the [111] NV group aligned with the bias field exhibits a splitting $\Delta$, while the remaining NVs in the other three crystal orientations are degenerate, contributing to the inner two peaks.
    Error bars represent 1$\sigma$ uncertainty from the photon shot noise.
    (b) Optical image of the thin film BSCCO exfoliated on a diamond surface.
    (c) Local magnetic field $B_z$ as a function of the applied field $H$ along a line-cut across the BSCCO [black dashed line in (b)] at 30~K.
    Compared to NV centers far away from BSCCO, those under BSCCO (gray region) experience smaller local magnetic fields due to the Meissner suppression.
    Near the edge of the BSCCO, $B_z$ gets enhanced due to the concentration of the magnetic field lines.
    Error bars represent the 1$\sigma$ uncertainty propagated from the fits to the ODMR spectrum.
    Right top: schematic of magnetic flux penetrating a superconductor in its vortex phase.
    Right bottom: schematic of magnetic flux being repelled by a superconductor in its Meissner phase.
    (d) A local field image of vortices at $T = 60$~K after  field-cooling the BSCCO flake at $H \approx 2$~G.
    The scale bar in the figure is 20~$\mu$m.
    }
    \label{fig:fig2}
\end{figure*}

\vspace{1em}
\emph{Experimental setup ---}
For our experiments, we study Bi$_2$Sr$_2$CaCu$_2$O$_{8+\delta}$ (BSCCO), a prominent type-II superconductor with a layered crystal structure and a high critical temperature $T_c$ exceeding the boiling point of liquid nitrogen~\cite{keimer2015quantum,zhou2021high,blatter1994vortices}.
This compound exhibits a rich and complex phase diagram, making it an ideal platform for investigating a variety of different types of low-energy dynamics~[Fig.~\ref{fig:fig1}(c)]. 
Bulk crystals of nearly optimally doped BSCCO are grown by the traveling floating zone method~\cite{wen2008large}, with $T_c\approx 90~$K measured via transport (see Methods).
To perform noise spectroscopy with NV centers, we exfoliate the BSCCO to a $ 200~$nm-thick film and transfer it directly onto a diamond sample containing an ensemble of NV centers in a thin layer, most of which are located $\lesssim 100$ nm beneath the diamond surface (see Supplementary Information)~[Fig.~\ref{fig:fig1}(b)].

Each NV center hosts a spin-1 electronic ground state, with Hamiltonian \cite{rovny2024nanoscale}: 
\begin{equation}
    \begin{aligned}
        \mathcal{H} &= D_\mathrm{gs} S_z^2 + \gamma_e B_z S_z \\
        &\quad + \Pi_x (S_y^2-S_x^2)
        + \Pi_y (S_x S_y + S_y S_x) \,,
    \end{aligned}
\end{equation}
where \{$S_x$, $S_y$, $S_z$\} are the spin-1 operators of the NV with the N-V axis defining the quantization axis ($\hat{z}$), $D_\mathrm{gs} = 2.87$~GHz is the zero-field splitting between $|m_s=0\rangle$ and $|m_s=\pm1\rangle$ sublevels, $\Pi_\perp = \sqrt{\Pi_x^2+\Pi_y^2} = 6.6$~MHz characterizes the coupling to local electric fields~\cite{mittiga2018imaging}, $B_z$ is the local magnetic field along the NV axis, and $\gamma_e$ is the electronic spin gyromagnetic ratio.

The NV energy levels can be probed via optically detected magnetic resonance (ODMR) spectroscopy, where one detects a decrease in NV fluorescence intensity when sweeping the frequency of an applied microwave through the resonances $|m_s=0\rangle \leftrightarrow |m_s=\pm1\rangle$.
In the experiment, we use a [111]-cut diamond crystal and apply a tunable magnetic field $H$ along the out-of-plane direction~[Fig.~\ref{fig:fig1}(b)].
In this geometry, the NV crystallographic group aligned with the applied field exhibits the largest splitting, $\Delta = \sqrt{(2\Pi_\perp)^2+(2\gamma_e B_z)^2}$ between $|m_s=\pm1\rangle$ sublevels, where $B_z$ includes both contributions from the applied field $H$ and the $B^{\mathrm{s}}_z$ generated from the superconducting sample ($B_z = H+B_z^\mathrm{s}$).
The other three NV groups are always observed to be degenerate.%

\vspace{1em}
\emph{Imaging the Meissner effect and flux vortices ---}
Using the NV centers as sensitive local magnetometers, we begin by determining the presence of superconductivity in BSCCO via the DC Meissner effect~\cite{schlussel2018wide, yip2019measuring, lesik2019magnetic, bhattacharyya2024imaging,nusran2018spatially,ho2025studying}.
After zero-field cooling the sample down to $10$~K (well below $T_c$), we perform an ODMR measurement with $H=0$ [Fig.~\ref{fig:fig2}(a)].
The resulting spectrum shows two resonances separated by the intrinsic splitting $2\Pi_{\perp}$.
When we apply a small field $H=15$~G, the ODMR spectrum remains largely unchanged, displaying only a slightly larger splitting, which suggests that the magnetic field generated by the sample cancels the applied one, $B_z^\mathrm{s} \approx -H$. 
Indeed, the extracted local field at the NV, $B_z = 2.5$~G $\ll H = 15$~G, corroborates the presence of a superconducting Meissner phase that expels the external field.
We emphasize that the observation of a near-complete cancellation of $H$ implies a close proximity of our NV sensors to the superconducting sample.

As we increase the temperature, the near-complete expulsion of the external field persists until $T\sim 40$~K. Beyond this point, the ODMR spectrum splits into a characteristic four resonances, indicating that the field is no longer fully canceled and that the system has transitioned into a vortex phase, allowing magnetic flux to penetrate the sample.
As the temperature further rises, the vortex density increases, raising the local field $B_z$ until it reaches the applied field $H$ at $T = T_c\approx 90$~K.
Interestingly, right below $T_c$, the local field $B_z$ is measured to be slightly greater than $H$, which may be attributed to a paramagnetic Meissner effect~\cite{geim1998paramagnetic}.

Leveraging the local nature of our sensor, we then investigate the spatial dependence of the Meissner suppression.
Performing ODMR spectroscopy along a line-cut across the BSCCO sample [black dashed line in Fig.~\ref{fig:fig2}(b)], we directly extract the local magnetic field $B_z$ as a function of spatial position at $T=30$~K [Fig.~\ref{fig:fig2}(c)].
Beneath the sample, the Meissner suppression is clearly observed at small applied field $H\lesssim20$~G.
However, when $H$ exceeds 25~G, which is on the same order as the independently characterized lower critical field $H_\mathrm{c1}$ (see Methods), the Meissner effect is suppressed, and $H$ is only partially expelled.
Crucially, the residual field also exhibits spatial variations across the sample, indicating that magnetic flux penetrates the BSCCO, forming superconducting vortices.
While the inter-vortex spacing for $H \gtrsim 25$~G  is too small to resolve them individually, we confirm the presence of frozen vortices in our sample by directly imaging their locations [Fig.~\ref{fig:fig2}(d)] using a widefield NV sensing microscope at $T = 60$~K, after field-cooling under a much smaller external field $H \approx 2$~G.

\begin{figure*}[t]
    \centering
    \includegraphics[width=0.9\textwidth]{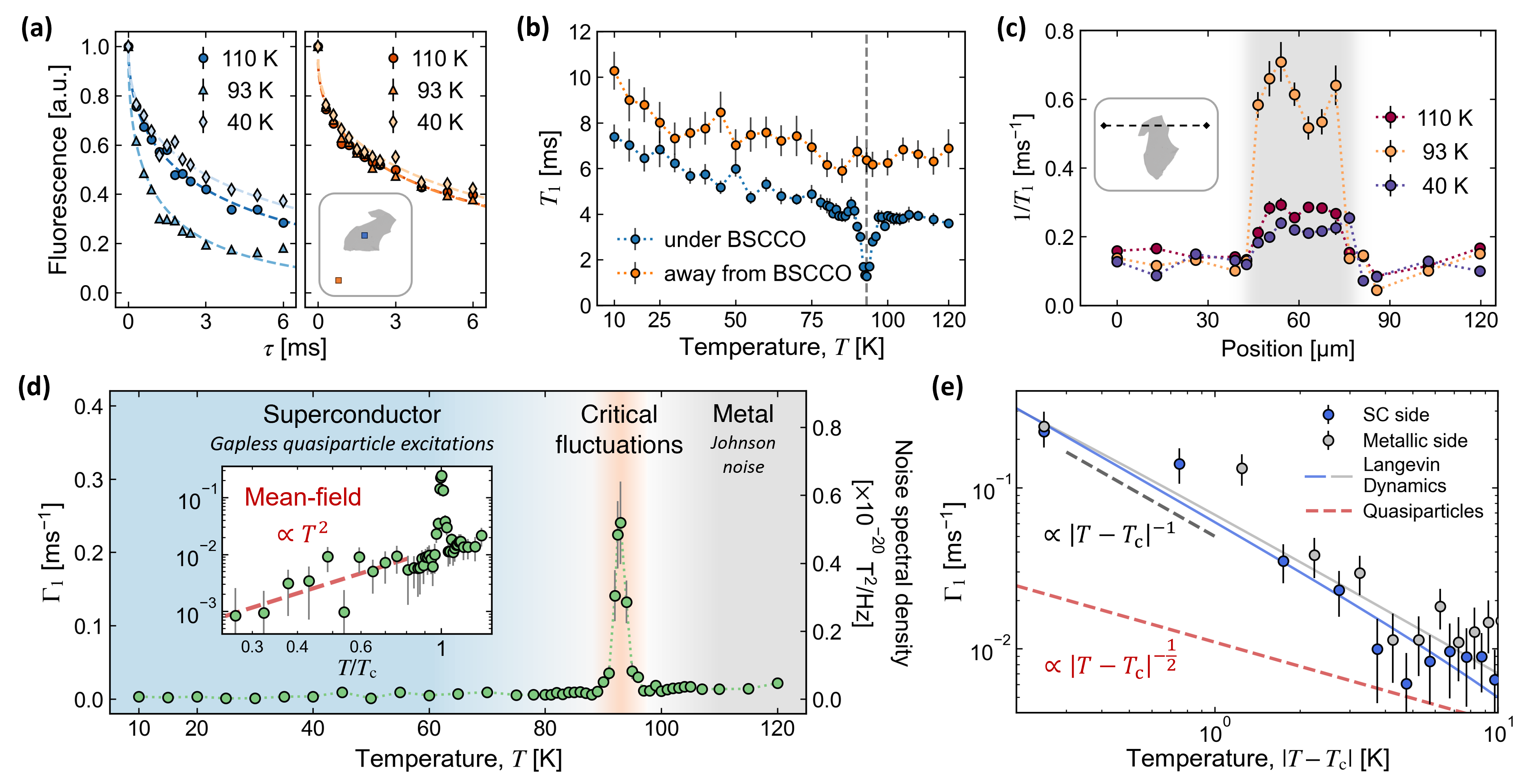}
    \caption{{\bf Zero-field superconducting fluctuations.}
    (a) $T_1$ decay profile for NV centers located under BSCCO (left, blue) and away from BSCCO (right, orange) at 110, 93, and 40~K.
    The dashed curves are stretched exponential fits to the data, with a fixed stretching exponent of 0.5.
    Error bars represent 1$\sigma$ uncertainty from the photon shot noise.
    Inset: Location of the measured NV centers with respect to the BSCCO sample.
    (b) The $T_1$ relaxation timescales as a function of temperature measured on the NV centers located under and away from BSCCO.
    (c) Spin relaxation rate, $1/T_1$, along a spatial line-cut across the BSCCO sample (see inset) at 110, 93, and 40~K. 
    The relaxation rate increases for NV centers under BSCCO (shadow region), particularly near the critical temperature $T_c$.
    (d) The BSCCO-induced decay rate of NVs, $\Gamma_1(T)$, as a function of temperature with $H = 0$.
    The corresponding magnitude of the noise spectral density is shown on the right axis.
    Inset: a power-law relationship at low temperatures, $\Gamma_1(T) \propto T^2$, is consistent with a mean-field approach accounting for nodal quasiparticles in $d$-wave superconductors.
    (e) $\Gamma_1(T)$ as a function of the absolute temperature deviation from the critical temperature, $|T - T_c|$, for both the metallic and superconducting (SC) sides of the transition.
    BCS mean-field quasiparticle dynamics (red dashed line) predicts $\Gamma_1(T) \propto |T - T_c|^{-x}$ with $x = 1/2$ on the superconducting side, and weak (non-diverging) temperature dependence on the metallic side ($x  = 0$).
    However, our experimental data are consistent with $x \approx 1$ (black dashed line) on both sides of the transition, suggesting that additional superconducting order parameter fluctuations are important.
    These data align closely with a theoretical model of critical Langevin dynamics of the order parameter (solid lines).
    Error bars indicate the $1\sigma$ uncertainty derived from the fits (b, c) and subsequent propagation (d, e).
    }
    \label{fig:fig3}
\end{figure*}

\vspace{1em}
\emph{Zero-field superconducting fluctuations ---}
Employing the full capabilities of the NV center, we go beyond conventional static magnetic field measurements and study superconducting fluctuations by performing $T_1$ relaxometry of the NV~[Fig.~\ref{fig:fig1}(b)].
To this end, we first optically polarize the NV spin state to $|m_s=0\rangle$ and subsequently measure the timescale $T_1$ over which the spin relaxes back to thermal equilibrium~[Fig.~\ref{fig:fig3}(a)].
Although spin-phonon interactions within the diamond itself lead to an intrinsic relaxation rate of NV centers, additional magnetic fluctuations from BSCCO, resonant with the NV's bare frequency ($D_\mathrm{gs} = 2.87$~GHz), can expedite the spin relaxation process and reduce the $T_1$ of NV.

This is indeed borne out by our data.
As shown in Figures~\ref{fig:fig3}(a) and (b), NV centers located beneath the BSCCO exhibit shorter $T_1$ compared to NV centers situated away from the BSCCO across the entire temperature range.
This trend is further highlighted in Figure~\ref{fig:fig3}(c), which shows the NV relaxation rate ($1/T_1$) as a function of a spatial line-cut across three distinct temperature regimes --- the superconducting phase, the critical region, and the metallic phase, with faster relaxation rates observed in all cases for NV centers beneath the BSCCO sample.

To quantitatively analyze the observed magnetic noise, we extract the BSCCO-induced decay rate, $\Gamma_1(T)$, by subtracting the intrinsic relaxation rate --- measured from NV centers located far from the BSCCO sample --- from the total relaxation rate of NV centers situated beneath the BSCCO [Fig.~\ref{fig:fig3}(d)].
At high temperatures, $T>T_c$, Johnson-Nyquist noise in the metallic phase determines the strength of the magnetic noise~\cite{kolkowitz2015probing,agarwal2017magnetic,hsieh2019imaging}. 
At very low temperatures, $T \ll T_c$, the magnetic noise is significantly suppressed.
Crucially, the form of this suppression offers insights into the nature of the superconducting pairing symmetry.
While in conventional s-wave superconductors $\Gamma_1(T)$ is exponentially suppressed at low temperatures due to a fully gapped Bogoliubov quasiparticle spectrum ~\cite{bardeen1957theory, bardeen1957microscopic,chatterjee2022single,dolgirev2022characterizing,kelly2024superconductivity,li2026observation}, the presence of nodal quasiparticles in $d$-wave layered superconductors implies a distinct power-law suppression of $\Gamma_1(T) \sim T^2$~\cite{chatterjee2022single,dolgirev2022characterizing}. 
Indeed, this expectation is consistent with our observations [Fig.~\ref{fig:fig3}(d), inset], corroborating the nodal $d$-wave nature of superconductivity in BSCCO~\cite{keimer2015quantum}.

The most striking feature of Fig.~\ref{fig:fig3}(d) is the sharp and symmetric peak in $\Gamma_1(T)$ near the critical temperature $T_c$, corresponding to a maximum noise spectral density of $10^{-20}$ Tesla$^2$/Hertz.
This peak arises from strong fluctuations in the superconducting order parameter, resulting in enhanced current fluctuations that act as a source of magnetic noise at the metal-superconductor phase transition.
Crucially, this enhancement is observed across the entire BSCCO sample [Fig.~\ref{fig:fig3}(c)].
Interestingly, near $T_c$, $\Gamma_1(T)$ begins to show marked deviations from mean-field expectations for $d$-wave superconductors. 
In particular, BCS mean-field theory predicts a power-law behavior of $\Gamma_1(T) \propto |T-T_c|^{-x}$ with $x=1/2$ near the transition on the superconducting side (up to logarithmic corrections)~\cite{dolgirev2022characterizing}, and a featureless $\Gamma_1(T)$ on the metallic side~\cite{agarwal2017magnetic}.
By contrast, the experimentally measured $\Gamma_1(T)$ at criticality is distinctly more singular, with $x \approx 1$ on both sides of the transition [Fig.~\ref{fig:fig3}(e)].
By fitting to power-law decays, we obtain $x = 1.10~\pm~0.09$ on the superconducting side and $x = 0.97~\pm~0.10$ on the metallic side.
Similar non-mean-field critical scaling behavior of $\Gamma_1(T)$, with a power law $x \approx 1$, is also observed in a second BSCCO sample exfoliated onto a different diamond sensor (see Supplementary Information).
This deviation arises because BCS theory neglects both amplitude and phase fluctuations of the superconducting order parameter $\psi$, which play a significant role near the critical point~\cite{emery1995importance, larkin2005theory, maki1968critical}. 
These effects are particularly important for thin-film superconductors, where fluctuations are expected to be stronger \cite{tinkham2004introduction}.
Consequently, a quantitative characterization of the divergent time scale associated with dissipative dynamics of the superconducting order parameter near $T_c$ remains an important open question, which we turn to next. 

To properly account for low-energy critical fluctuations, we consider the following phenomenological Langevin dynamics of the superconducting order parameter $\psi(\bm{\rho},t)$:
\begin{equation}
    \begin{aligned}
        \partial_t \psi(\bm{\rho},t)
        &= -\gamma \frac{\delta F}{\delta \psi^*(\bm{\rho},t)}
        + \eta(\bm{\rho},t), \\
        \text{ with } F[\psi]
        &= \int d\bm{\rho} \,
        \left[K |\bm{\nabla}\psi|^2 + r(T) |\psi|^2
        + \frac{u}{2} |\psi|^4 \right],
    \end{aligned}
\end{equation}
where $F[\psi]$ is the Ginzburg--Landau free energy functional \cite{lifshitz1980statistical,cyrot1973ginzburg,schuller2006time,tinkham2004introduction}, $\eta$ is local white noise that captures the effect of coarse-graining, and $\gamma$ sets the rate of relaxation towards equilibrium \cite{hohenberg1977theory}.
Crucially, by setting $r(T) \propto T - T_c$ near the critical point, we can analyze fluctuations of the order parameter around the minima of free energy associated with both phases: $\langle \psi \rangle = 0$ for $T > T_c$ (metallic) and $\langle |\psi| \rangle = \psi_0$ for $T < T_c$ (superconducting). 
These order parameter fluctuations manifest themselves as current fluctuations in the BSCCO sample, which induce magnetic noise at the NV's frequency $D_\mathrm{gs}$, enhancing its relaxation rate $\Gamma_1(T)$.
This low-frequency fluctuation enhancement is directly connected to the phenomena of critical slowing down near the phase transition; the noise spectral density at low frequencies scales with the scattering timescale, which diverges as one approaches the critical point.
For weak-coupling superconductors, this divergent timescale is quantified by $\tau_{\rm GL} = \pi\hbar/(8k_B|T - T_c|)$ near $T_c$~\cite{larkin2005theory,schmid1966time}.

We find that our theoretical model (see Methods) can faithfully reproduce the divergence of $\Gamma_1(T)$ on both sides of the transition [Fig.~\ref{fig:fig3}(e)].
At the same time, our approach enables us to quantitatively extract the value of the scattering timescale $\tau^{\text{fit}}$ by fitting our model to the experimental data on both the superconducting [sc] and metallic [m] sides of the transition.
Remarkably, the extracted value for the scattering timescale, given by $\tau^{\rm fit}_{\rm sc[m]} / \tau_{\rm GL} \approx 0.3[2.1] $, are comparable to the analytical weak-coupling results, $\tau^{\rm wc}_{\rm sc[m]} / \tau_{\rm GL} = 0.5[1]$ \cite{larkin2005theory,schmid1966time,cheng2026qubit}.
Understanding the broad agreement between the weak-coupling prediction and the expected strong-coupling nature of BSCCO superconductivity remains an open question~\cite{lee2006doping}. 

The agreement between our theoretical model and the experimental data suggests that the critical behavior is properly captured by our Langevin dynamics approach.
Indeed, the resulting values of the associated critical exponents, $\nu=1/2$ and $z=2$~\cite{hohenberg1977theory}, directly reproduce the aforementioned divergence of the scattering timescale $\tau$ via the scaling behavior of the correlation length $\xi$: $\tau \propto \xi^z \propto  |T - T_c|^{-z\nu}$.
Further, the approximately symmetric decay of $\Gamma_1(T)$ on both sides of its peak at $T_c$ indicates that vortex-antivortex fluctuations, which are predicted to lead to a highly asymmetric $\Gamma_1(T)$~\cite{curtis2024probing}, are relatively less important in determining the critical behavior of the BSCCO flake.

\begin{figure*}[t]
    \centering
    \includegraphics[width=0.9\textwidth]{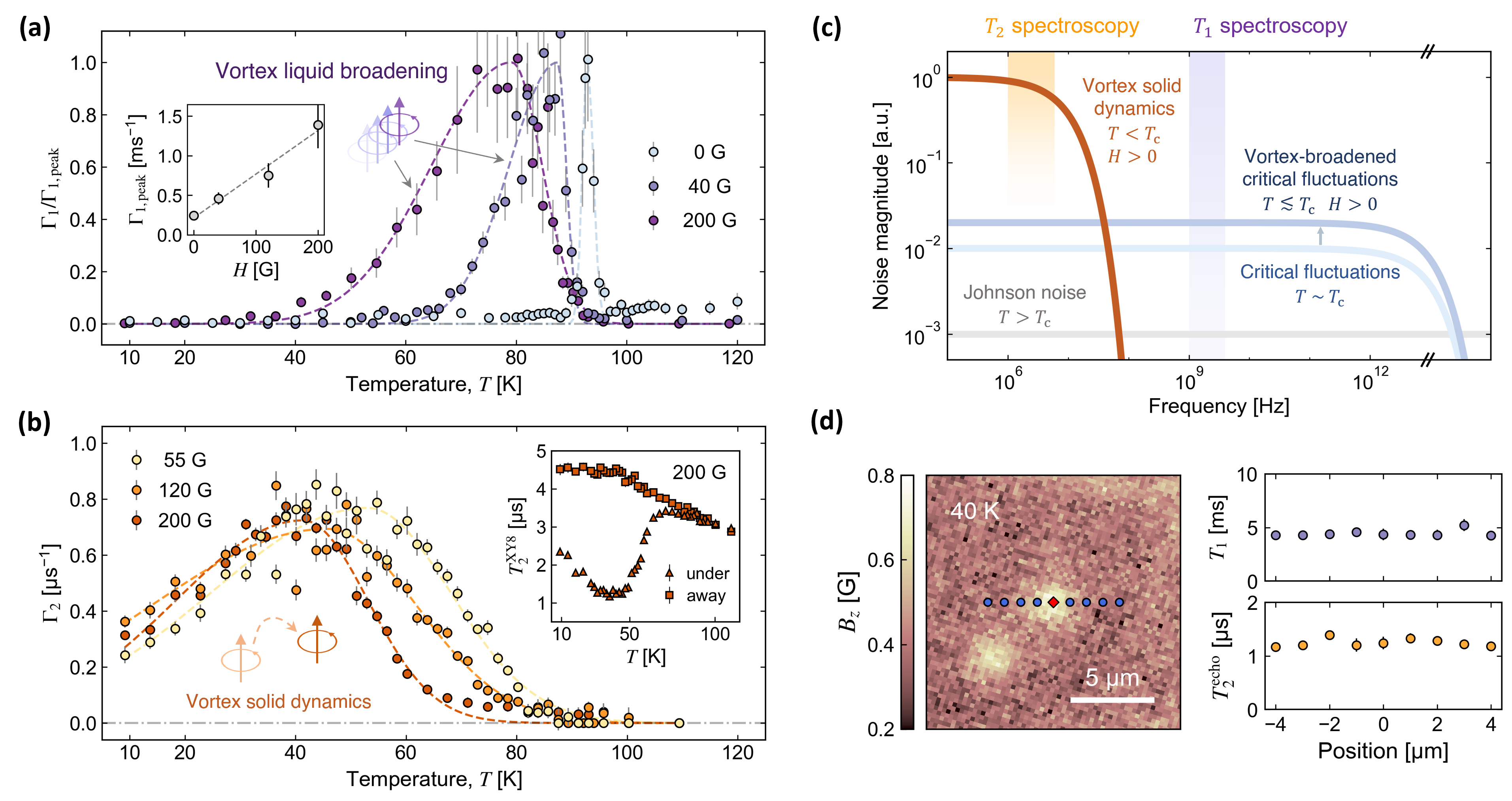}
    \caption{{\bf In-field criticality and vortex dynamics.}
    (a) BSCCO-induced relaxation rate, $\Gamma_1(T)$ (normalized), as a function of temperature at $H = 0, 40, 200$~G.
    The dashed curves represent phenomenological fits to the data using an asymmetric Gaussian function.
    Inset: peak magnitude of $\Gamma_1(T)$ as a function of applied field, which is consistent with a linear dependence on the field strength $H$ (gray line).
    Peak values correspond to data points that are closest to the fitted peak temperatures.
    (b) BSCCO-induced decoherence rate, $\Gamma_2(T)$, as a function of temperature at $H = 55, 120, 200$~G, fitted with asymmetric Gaussian functions (dashed curves).
    Inset: The measured $T^\mathrm{XY8}_2$ decoherence timescale as a function of temperature for NVs under and away from BSCCO at $H = 200$~G.
    (c) Schematic of the noise spectrum arising from different dynamical processes in BSCCO in the context of our two probing techniques, NV relaxation ($T_1$) and decoherence ($T_2$) noise spectroscopy.
    While less sensitive than $T_1$ spectroscopy, $T_2$ spectroscopy is able to access fluctuations at much lower frequencies and observe vortex solid dynamics.
    (d) Left: two isolated pinned vortices imaged in BSCCO under $H \approx 0.5$~G at 40~K.
    The dots mark the positions for line-cut measurements.
    Right: measured relaxation time $T_1$ and decoherence time $T^\mathrm{echo}_2$ of NV centers at different spatial positions across a vortex centered at zero.
    Error bars represent the $1\sigma$ uncertainty from fitting the $T_1$ and $T_2$ decay curves.
    }
    \label{fig:fig4}
\end{figure*}

\vspace{1em}
\emph{In-field criticality and vortex dynamics ---}
Next, we turn on the applied external field and explore the in-field critical fluctuations at $H=40$~G and $H=200$~G.
Intriguingly, the measured $\Gamma_1(T)$ differs from the zero-field case in two key ways [Fig.~\ref{fig:fig4}(a)]: 
(i) The peak position of the noise is slightly shifted towards lower temperatures. 
(ii) The peak becomes broad and highly asymmetric; while the noise signal remains sharp when approaching criticality from the metallic side, it increases much more gradually and smoothly from the superconducting side.
The former is expected, as a perpendicular field hinders phase coherence and lowers the critical temperature $T_{c}$. 
The latter can be attributed to two different factors.
On the one hand, an external magnetic field broadens the temperature range where critical fluctuations are important, as described by two-dimensional XY critical dynamics~\cite{beasley1979possibility,halperin1979resistive,curtis2024probing}. 
On the other hand, below $T_c$, an intermediate vortex liquid phase is expected to form, where the local motion of the superconducting vortices (which trap magnetic flux quanta) can generate an additional fluctuating magnetic field at the NV~\cite{blatter1994vortices,fisher1991thermal,nelson1988vortex}. 

Further evidence for the latter can be gleaned from the scaling of the peak magnitude of $\Gamma_1$ with the magnetic field $H$ [inset, Fig.~\ref{fig:fig4}(a)], where the density of vortices $n_v$ increases linearly with increasing field strength.
Within a model of Langevin dynamics of mobile vortices (see Supplementary Information), the magnetic noise generated by individual vortices adds incoherently when their motions are weakly correlated.
Consequently, the measured relaxation rate due to vortex-motion induced magnetic noise scales linearly with the vortex density, i.e., $\Gamma_1(H) \propto n_v \propto H$, in agreement with our observations [inset, Fig.~\ref{fig:fig4}(a)].

Our NV-based quantum noise spectroscopy provides a direct route to quantify an effective vortex pinning scale.
By fitting the temperature dependence of $\Gamma_1$ to our theoretical model, we extract an effective spring constant $K(T)$ for the restoring force on each vortex, arising from local disorder-pinning and inter-vortex repulsion.
We find $K(T \approx 70~\mathrm{K}) \sim 10^4~\mathrm{N}\mathrm{m}^{-2}$, comparable to optimally doped La$_{2-x}$Ce$_{x}$CuO$_4$~\cite{zuev2003vortex} and YBa$_2$Cu$_3$O$_{7-\delta}$~\cite{pesetski2000experimental}.
Further, $K(T)$ decreases as the temperature is increased, indicating weaker pinning and enhanced vortex mobility.

As the temperature further decreases, the vortices either freeze into a lattice or become pinned by defects, resulting in a decrease of $\Gamma_1(T)$. 
This conclusion is consistent with our measurement of noise along a line-cut across a pinned vortex shown in Fig.~\ref{fig:fig4}(d).
We purposefully chose to measure this noise at a much smaller magnetic field ($H \approx 0.5$ G), where we can resolve individual vortices via static field measurements, and at a low enough temperature where all vortices are expected to be strongly pinned: thereby avoiding additional contributions to the noise due to mobile vortices that cannot be resolved by static imaging. 
$T_1$ does not change appreciably across the vortex, indicating that strongly pinned vortices do not exhibit dynamics in the GHz frequency range.

\vspace{1em}
\emph{$T_2$ noise spectroscopy of slow vortex dynamics ---}
At sufficiently low temperatures, the system is deep in the superconducting phase and the vortex liquid is expected to transition into a solid state.
Nevertheless, with $H>H_{c1}$, BSCCO can still exhibit complex vortex dynamics, such as thermally assisted flux jumps between different pinning centers, or the depinning of vortices to exit the superconducting sample, leading to fluctuating magnetic signals from different physical processes.
However, the measured $\Gamma_1(T)$ in the low-temperature regime does not display any features, suggesting that these processes operate at even lower frequencies, and therefore elude detection by $T_1$ spectroscopy sensitive only to GHz-range fluctuations~[Fig.~\ref{fig:fig4}(c)].
 
To investigate these dynamics, we perform NV decoherence ($T_2$) noise spectroscopy to capture the magnetic noise spectrum in the MHz frequency range~\cite{machado2023quantum,xue2026magnon,ziffer2024quantum,li2025critical}.
Specifically, after preparing the NV center into a quantum superposition state, $(|m_s=0\rangle+|m_s=-1\rangle)/\sqrt{2}$, we apply a dynamical decoupling sequence (XY-8, see Supplementary Information) to measure the decoherence rate induced by BSCCO, $\Gamma_2(T)$, as a function of both temperature and applied field $H$ [Fig.~\ref{fig:fig4}(b)].
Remarkably, $\Gamma_2(T)$ exhibits a broad noise spectrum with several prominent features. 
(i) With increasing field $H$, there is a reduction of the onset temperature for $\Gamma_2$-type noise, and this onset occurs for temperatures below the peak position of $\Gamma_1(T)$.
(ii) For different $H$, $\Gamma_2(T)$ reaches its maximum value at a much lower temperature $T^*$ than $\Gamma_1(T)$; however, the applied field $H$ remains larger than the independently measured $H_{c1}$ of the sample at $T^*$ (see Methods), indicating that vortices are still present in the sample.
(iii) The maximum value of $\Gamma_2(T)$ is insensitive to the strength of the external field $H$.

To understand these observations, we must rule out some of the previously studied mechanisms.
In particular, we begin by noting that the decoherence time $T_2$ on and off BSCCO is nearly identical in both the metallic phase and near criticality [inset of Fig.~\ref{fig:fig4}(b)]. 
This observation indicates that $T_2$ noise spectroscopy is less sensitive than $T_1$ in detecting both Johnson-Nyquist noise and critical current fluctuations.
This difference in sensitivity can be attributed to the separation between the intrinsic depolarization and dephasing timescales of the NV.
More specifically, the NV's $T_1$ is around three orders of magnitude larger than its $T_2$; a significantly stronger noise amplitude is required to induce a measurable change of NV $T_2$~(Methods).

At the same time, $T_2$ measured across a line-cut intersecting a pinned vortex shows no position dependence, indicating that the noise amplitude from the core of a static vortex in the MHz frequency range is below our measurement threshold [Fig.~\ref{fig:fig4}(d)].
Our measurement therefore suggests that $T_2$ is less sensitive to static vortices in BSCCO films, and most sensitive to vortex motion within the vortex solid phase~\cite{fisher1989vortex,fisher1991thermal}.
More specifically, the noise magnitude implies that it likely originates from thermal fluctuations, which can cause positional jumps of fluxes or flux bundles between different weak pinning centers, resulting in sudden large changes in the local magnetic field. 
These dynamics are relatively slow, extending only up to the MHz range. 
Thus, they are not reflected in $\Gamma_1$, but can be directly captured by $\Gamma_2$ [Fig.~\ref{fig:fig4}(c)].

These observations collectively lead to the following plausible scenario that accounts for the temperature dependence of both $\Gamma_1$ and $\Gamma_2$ due to vortex motion.
As the temperature is initially lowered, sample inhomogeneities lead to disorder-pinning of an increasing fraction of mobile vortices.
The slow jumps of these pinned vortices between different pinning centers are detected by $\Gamma_2$, leading to an increase in $\Gamma_2(T)$ with decreasing $T$. 
This increase in $\Gamma_2(T)$ is concomitant with a reduction of $\Gamma_1(T)$, which detects lower mobility of vortices at the GHz scale due to stronger pinning at lower temperatures, and is insensitive to flux-jump motion at the MHz scale.
Only at very low temperatures, approximately corresponding to the Meissner phase for bulk BSCCO, does such flux-jump motion begin to freeze out, leading to a decrease in $\Gamma_2(T)$ below $40$~K.

\vspace{1em}
\emph{Outlook ---}
For the first time, we demonstrate the use of NV centers as nano-scale quantum sensors of superconducting criticality and vortex dynamics.
The nano-scale proximity and high sensitivity of our NV sensors, combined with their operational frequency range spanning both MHz and GHz regimes and sub-micron spatial resolution, enable the measurement of a wide swath of dynamical phenomena beyond the capabilities of conventional techniques [Fig.~\ref{fig:fig1}(a)]. 

Looking forward, our work opens the door to several intriguing directions. 
First, while our current studies involve ensemble NV centers with a spread of sensor-sample distances, an immediate future direction is to precisely adjust the distance $d$ between the quantum sensor and the target material.
This can be achieved by using a single NV center on a scanning tip~\cite{maletinsky2012robust, degen2008scanning, pelliccione2016scanned}, or by creating spin sensors in a given layer of Van der Waals materials such as hexagonal boron nitride~\cite{gottscholl2020initialization,gong2023coherent,stern2022room,gong2024isotope}.
By tuning the distance, we can adjust the \emph{momentum} filter function to enable controlled probing over a variety of length scales (from nanometers to micrometers)~\cite{machado2023quantum}, which is crucial for imaging inhomogeneities in superconductors~\cite{bhattacharyya2024imaging, dailledouze2025imaging}.

Moreover, our measurement of the critical slowing-down time scale $\tau$ places strong constraints on theoretical models of the transition and provides new opportunities to study the mechanisms for order parameter relaxation near criticality.
In fact, a reasonable agreement with the weak-coupling result $\tau = 0.5 \, \tau_{\rm GL}$ on the superconducting side~\cite{schmid1966time}, but a stronger deviation on the metallic side~\cite{larkin2005theory}, indicates the necessity for further microscopic investigations of critical dynamics in unconventional superconductors. 

Finally, our technique is not limited to BSCCO, and can be readily extended to a variety of superconducting materials. For instance, certain superconductors such as hydrides~\cite{pickard2020superconducting, drozdov2015conventional} and nickelates~\cite{sun2023signatures, wen2025imaging} undergo superconducting transitions under high pressure. 
Thus, moving beyond temperature-driven transitions, the integration of NV centers into high-pressure diamond anvil cells offers a pathway to study how pressure influences critical fluctuations and vortex dynamics~\cite{hsieh2019imaging,lesik2019magnetic, yip2019measuring,bhattacharyya2024imaging,dailledouze2025imaging}.

\vspace{1em}
\emph{Acknowledgments ---}
We gratefully acknowledge discussions with Pavel Dolgirev, Chris Laumann, Paul Ching-Wu Chu, Liangzi Deng, Andrew Mounce, Ania Jayich, Peng Cai, Panyu Hou, Srinivas Mandyam, Zhipan Wang.
This work is supported by NSF 2514391, NSF NRT LinQ 2152221, NSF ExpandQISE 2328837, and the Center for Quantum Leaps at Washington University.
O.\,K.\,D. and F.\,M. acknowledge support from the NSF through a grant for ITAMP at Harvard University. 
The work at Brookhaven National Laboratory is supported by the US Department of Energy, Office of Basic Energy Sciences, contract No.~DOE-sc0012704.
E.\,A.\,H. and X.\,D. acknowledge support from the Gordon and Betty Moore Foundation, grant DOI~10.37807/gbmf11560.
S.\,R., Q.\,X., and Z.\,R. acknowledge support under National Science Foundation (NSF) Division of Materials Research Award DMR-2236528. 
S.\,C. acknowledges support from a PQI Community Collaboration Award.

\putbib[references]

\end{bibunit}


\clearpage
\onecolumngrid

\bookmarksetup{startatroot}
\bookmarksetupnext{level=section}
\majorsection{Methods}{methods}

\methodsnumbering

\setbiblinksuffix{-methods}
\begin{bibunit}[apsrev4-2-etal]

\methodsection{Characterization of BSCCO}

\methodsubsection{Resistance}

\begin{figure*}[h!]
    \centering
    \includegraphics[width=0.95\textwidth]{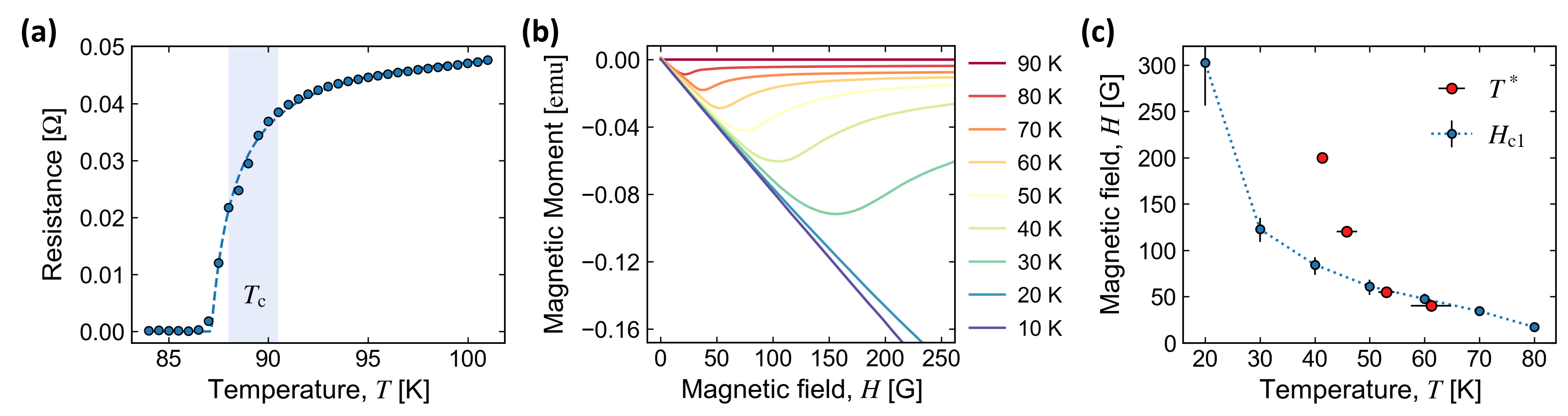}
    \caption{{\bf Characterization of BSCCO samples.}
    (a) Electrical resistance as a function of temperature.
    The transition temperature $T_c$ is indicated by a shaded region.
    (b) Magnetic moment versus magnetic field ($M$--$H$) curve at different temperatures.
    (c) Extracted lower critical field, $H_\mathrm{c1}$, as a function of temperature.
    The $T^*$ marks the peak positions in $T_2$ spectroscopy [see Fig.~4(b)] at applied field $H = 40, 55, 120, 200$~G, with error bars determined from asymmetric Gaussian fittings.
    }
    \label{fig:fig_char}
\end{figure*}

Electrical transport measurements are performed on a bulk BSCCO sample, which is grown from the same batch as the exfoliated sample employed in NV experiments, using a Quantum Design Physical Property Measurement System (PPMS).
An electric current is applied along the $ab$ plane, and the resistance is measured in the same direction using the four-point-probe method.
As shown in Figure~\ref{fig:fig_char}(a), the resistance drops to zero below the superconducting transition temperature at $T_c \approx 90$~K.

\methodsubsection{Lower critical field}

Lower critical field measurements are performed on a bulk BSCCO sample using a Quantum Design PPMS equipped with the Vibrating Sample Magnetometer (VSM) option. 
The lower critical field of BSCCO is extracted from the Magnetization--Field ($M$--$H$) curves, see Figure~\ref{fig:fig_char}(b).
The magnetization here is replaced by the magnetic moment since the sample volume stays unchanged.
For each temperature, the initial diamagnetic region is fitted to a linear line, and the R-square value is calculated.
By gradually expanding the fitting region, we cap the endpoint at where R-square starts to drop below a threshold ($r^2 < 0.998$), and define the cap field as the lower critical field $H_\mathrm{c1}$ at that temperature.
As a result, the temperature dependence of the lower critical field, $H_\mathrm{c1}(T)$, is extracted and summarized in Figure~\ref{fig:fig_char}(c).
The error bars are determined by setting an upper bound (0.999) and a lower bound (0.996) of the threshold (0.998).

\methodsection{Sensitivity of $T_1$ and $T_2$ spectroscopy}

For small noise, the coherence signal behaves as:
\begin{equation}
    p(t) = C_0 \exp\left\{
    -\left(\frac{t}{T_{\text{int}}}\right)^\alpha
    - \left[t \Gamma\right] \right\}
    = e^{-\chi(t)} e^{-t\Gamma} \,,
\end{equation}
where $e^{-\chi(t)}$ encodes the intrinsic decay (either depolarization or decoherence) that the NV experiences with decay time scale of $T_{\text{int}}$: $e^{-\chi(T_{\text{int}})} = e^{-1}$; and $\Gamma$ encodes the additional decay of the NV due to fluctuations in the sample.

Following the discussion in \cite{degen2017quantum}, a signal is measurable whenever the signal-to-noise ratio (SNR) is unity.
This enables us to compute the smallest measurable signal given the experimental conditions:
\begin{equation}
    \text{SNR} = \delta \Gamma
    \left[ |-t e^{-t\Gamma}|\right|_{\Gamma=0}
    e^{-\chi(t)} 2 C \frac{\sqrt{T}}{\sqrt{t+t_m}}
    \propto \delta \Gamma
    \frac{t e^{-\chi(t)}}{\sqrt{t+t_m}} \ge 1 \,,
\end{equation}
where $C$ encodes the readout efficiency of the protocol, $T$ is the total integration time, $t$ is the duration of the measurement, and $t_m$ is the additional time required for initialization and readout.
Owing to the exponential-like intrinsic decay of the qubit's coherence, the optimal value of the experiment's duration is the one that maximizes $t e^{-\chi(t)}$, which occurs when $t\sim T_{\text{int}}$.
This implies that the smallest value of $\Gamma$ that is measurable scales as:
\begin{equation}
    \delta \Gamma  \gtrsim  \frac{e \sqrt{T_{\text{int}} + t_m}}{T_\text{int}} \sim
    \begin{cases}
        \frac{1}{T_{\text{int}}}, & T_{\text{int}} \ll t_m\\
        \frac{1}{\sqrt{T_{\text{int}}}}, & T_{\text{int}} \gg t_m
    \end{cases}
    \,.
\end{equation}
As a result, as the intrinsic decay time increases, one is able to measure a smaller signal for the same integration time --- there is an improvement in sensitivity.

\methodsection{Theoretical Analysis}

Here, we provide the theoretical methods used to model the experimental data, in particular, focusing on the $T_1$ measurements.
At zero applied magnetic field ($H = 0$), we employ a time-dependent Ginzburg--Landau formalism to model the order-parameter dynamics around the critical point and find a qualitative match with experimental observations of the depolarization rate $\Gamma_1(T)$, both on the superconducting and metallic side of the phase transition.
When $H \neq 0$, we show that $\Gamma_1(T, H)$ originating from weakly correlated motion of vortices scales linearly with the magnetic field $H$, as observed.
Further, in the Supplementary Information, we use Langevin dynamics of vortices to fit the experimental data and extract an effective spring constant $K(T)$ that controls the restoring force on vortices displaced from their equilibrium positions.

The depolarization of the qubit occurs predominantly due to fluctuations of the local magnetic field $\B(\r_{\rm NV},t)$ at the NV location.
Using Fermi's golden rule, the experimentally measured relaxation rate is related to the transverse magnetic noise spectral density $\mathcal{N}_{\rm T}(\Omega)$ by \cite{kolkowitz2015probing}:
\begin{equation}
    \Gamma_1 \equiv \frac{1}{T_1} = \frac32\gamma_e^2\mathcal{N}_{\rm T}(\Omega) \,,
    \label{Eq:DepolarizationRate}
\end{equation}
where $\gamma_e=g \mu_B / \hbar$ is the electronic spin gyromagnetic ratio, with $g \approx 2$ being the NV ground-state Landé $g$-factor, $\mu_B$ the Bohr magneton, and $\hbar$ the reduced Planck constant.
Hence, our measurement of $\Gamma_1$ directly gives us access to the BSCCO-induced magnetic noise at the NV operational angular frequency $\Omega$ via Eq.~\eqref{Eq:DepolarizationRate}, as shown in Fig.~\ref{fig:fig3}(d) in the main text.

To model the magnetic noise experienced by the NV centers, we treat BSCCO as a set of two-dimensional active conducting CuO$_2$ bilayers that are weakly coupled.
To account for the 200 nm thickness of the BSCCO flake, we consider $60$ unit cells in the $c$ ($\hat{\z}$) direction, with $2$ active CuO$_2$ bilayers per unit cell.
Although the NV centers exhibit a distribution of depths (see Supplementary Information), we use a representative depth of $z_0=40$ nm in our calculations for simplicity.
Hence, we sum the magnetic noise from two active CuO$_2$ layers at $z_{\rm NV} = 40 + 1.55 \, n$ nm, with $n \in [0,120]$ (implying a total of 120 independent active bilayers).
We note that owing to the summation over all BSCCO layers, our results are qualitatively robust to the precise value of the average NV-surface distance, as long as it remains smaller than the thickness $d$ of the BSCCO flake.

\methodsubsection{Noise from critical fluctuations}
We begin by computing the transverse magnetic noise from a single  BSCCO bilayer.
Following the formalism set up in Refs.~\cite{dolgirev2022characterizing,chatterjee2022single}, the transverse magnetic noise experienced by the NV center at a distance $z_{\rm NV}$ from the sample is given as,
\begin{equation}
    \begin{aligned}
        \mathcal{N}_{\rm T}(\Omega) &= \frac{\mu_0 k_B T}{16 \pi z_{\rm NV}^3 \Omega}
        \int_0^{\infty} dx \, x^2 e^{-x}
        \mathrm{Im} \left[ r_s \left( \frac{x}{2 z_{\rm NV}},\Omega \right) \right] \,, \\
        \text{where } r_s(\q,\Omega) &= - \left(1 + \frac{2 i q}{\mu_0 \Omega ~ \sigma_{\rm
        T}(\q,\Omega)} \right)^{-1} \simeq
        \frac{i \mu_0 \Omega \, \sigma_{\rm T}(\q, \Omega)}{2q} \,.
    \end{aligned}
    \label{eq:NoiseEq}
\end{equation}
Here, $r_s$ denotes the reflection coefficient of $s$-polarized waves, which is determined by the transverse nonlocal conductivity $\sigma_{\rm T}(\q,\Omega)$.
The approximation requires $\mu_0\Omega z_{\rm NV}|\sigma_{\rm T}|\ll1$, as the characteristic wavevector $q\sim(2z_{\rm NV})^{-1}$. This gives $z_{\rm NV}\ll\Lambda/2$ for thin-film superconductors in the London limit, where Pearl length $\Lambda=2\lambda^2/d$ \cite{pearl1964current}, and $z_{\rm NV}\ll(\mu_0\Omega\sigma_n)^{-1}$ for normal metals, with a finite low-frequency sheet conductivity $\sigma_n$. Both conditions are readily satisfied for nanometer-scale NV depths because the relevant screening lengths, $\Lambda/2$ and \((\mu_0\Omega\sigma_n)^{-1}\), are on micrometer scales or larger.
Therefore, a primary theoretical objective is to calculate the transverse nonlocal conductivity $\sigma_{\rm T}(\q,\Omega)$, from which we may find the magnetic noise and subsequently the depolarization rate.

The transverse nonlocal conductivity $\sigma_{\rm T}(\q, \Omega)$ is determined from the transverse current-current correlator by the Kubo formula \cite{kubo1966fluctuation}.
In particular, since the NV frequency at which we probe critical fluctuations is much smaller than any other energy scale in the problem, we focus on the classical limit ($\hbar\Omega \ll k_B T$) of $\sigma_{\rm T}(\q,\Omega)$,
\begin{equation}
    \sigma_{\rm T}(\q,\Omega=0) = (k_B T)^{-1} \int_0^\infty \, dt \,
    \langle \J^{\rm T}_{\q}(t)\cdot \J^{\rm T}_{-\q}(0) \rangle \,.
    \label{Eq:TransverseConductivity}
\end{equation}
Here, $\J^{\rm T}$ denotes the transverse current, defined as, $\J^{\rm T}_{\q}=\J_{\q}-\left(\q\cdot \J_{\q}/\q^2\right)\q$, with the current operator $\J$ given by, 
$\J(\bm{\rho})=\left[\hbar e^* /\left(2 i m^*\right)\right]\big[\psi^*\bm\nabla \psi-(\bm\nabla \psi^*)\psi\big]$, where $e^*$ and $m^*$ denote the effective charge and mass of the charge carriers.

To determine the transverse current-current correlator, we must first determine the order-parameter correlation functions.
To this end, we model the dynamics of the order parameter $\psi$: we adopt a time-dependent Ginzburg--Landau theory approach \cite{cyrot1973ginzburg,lifshitz1980statistical, ginzburg1950theory,schuller2006time,tinkham2004introduction}. 
The Langevin equations for $\psi$ are given by
\begin{equation}
    \partial_{t}\psi(\bm{\rho}, t) = -\gamma \frac{\delta F}{\delta\psi^{*}(\bm{\rho},
    t)}+\zeta_\psi(\bm{\rho}, t) \,,
    \label{eq:Langevin}
\end{equation}
where $\psi(\bm{\rho}, t)$ denotes the order parameter, $\zeta_\psi(\bm{\rho}, t)$ is the zero-mean Gaussian noise that captures the effect of the high-energy modes on low-energy, long-wavelength order-parameter fluctuations, with correlations given by
\begin{equation}
    \langle \zeta_\psi(\bm{\rho}, t)\zeta_\psi^*(\bm{\rho}', t') \rangle = 2 \gamma k_B T \,
    \delta(t-t') \delta^{(2)}(\bm{\rho}-\bm{\rho}') \,,
\end{equation}
and $F$ is the $U(1)$-symmetric Ginzburg--Landau free-energy functional \cite{lifshitz1980statistical} given by 
\begin{equation}
    F=\int_{\bm{\rho}} \left(r|\psi|^{2}+K|\bm\nabla\psi|^{2}+\frac{u}{2}|\psi|^{4}\right) \,,
\end{equation}
where $K = \hbar^2/{(2m^*)}$.
The parameter $\gamma$ in Eq.~\eqref{eq:Langevin} depends on the coupling with the bath generating the dissipative dynamics, and can be calculated microscopically in certain weak-coupling limits \cite{larkin2005theory,schmid1966time,schuller2006time,tinkham2004introduction}.

In what follows, we determine the transverse nonlocal conductivity for both the metallic and the superconducting side using the time-dependent Ginzburg--Landau theory, and find the corresponding NV center's sample-induced depolarization rate $\Gamma_1(T)$.
We find that near criticality, the depolarization rate scales as $(T-T_c)^{-1}$ in accordance with the experiment. 
Further, as discussed in the main text, the extracted scattering timescales $\tau^{\rm fit}_{\rm m[sc]}$ are comparable to the weak-coupling Aslamazov-Larkin analysis.

\textbf{Metallic Side}:
On the metallic side of the transition, the minimum of the free energy $F$ sits at $\psi$ = 0. 
Ergo, we simplify Eq.~\eqref{eq:Langevin} further by ignoring the quartic term in $F$, upon which we obtain a linearized time-dependent GL equation:
\begin{equation}
    \frac{\partial \psi}{\partial t}
    = -\gamma(-K \nabla^2+r)\psi + \zeta_\psi
    = -\gamma r\left(1-\frac{K}{r}\nabla^2\right)\psi
    \equiv -\frac{1}{\tau_{\rm m}}
    (1-\xi_{\rm m}^2\nabla^2)\psi + \zeta_\psi \,,
    \label{eq:metalGzL}
\end{equation}
where we have defined $(\tau_{\rm m})^{-1} = \gamma r \propto T - T_c$ as the temperature-dependent relaxation rate of the uniform mode of the order parameter $\psi_{\k = 0}$, which shows critical slowing down near criticality. 
Further, we have also used the fact that the mean-field superconducting correlation length $\xi_{\rm m}(T)$ can be extracted from the static GL free energy $F$ as $\xi_{\rm m}^2(T) = K/r(T)$.
Since $r(T) \propto T - T_c$, the correlation length $\xi_{\mathrm{m}}(T)$ diverges near the critical temperature as $\xi_{\mathrm{m}} \propto \big[T_c/(T-T_c)\big]^{1/2}$, corresponding to a critical exponent of $\nu = 1/2$~\cite{tinkham2004introduction}.
Further, as the dynamical critical exponent $z$ is defined via the scaling form $\tau_{\rm m} \propto \xi_{\rm m}^z$, the divergence of $\tau_{\rm m} \propto (T - T_c)^{-1}$, as discussed earlier, implies $z = 2$. 
The precise proportionality constant for the divergence of $\tau_{\rm m}$ may be obtained using a weak-coupling Aslamazov-Larkin analysis \cite{cyrot1973ginzburg,larkin2005theory,schmid1966time}, which leads to $\tau_{\rm m}^{\text{wc}} = \tau_{\rm GL} = \pi \hbar/[8 k_B (T - T_c)]$. 
Since BSCCO is not necessarily in the weak-coupling regime, we will treat $\tau_{\rm m}$ as a phenomenological parameter, and extract the ratio $\tau_{\rm m}/\tau_{\rm GL}$ by fitting to the experimental data. 

Using Eq.~\eqref{eq:metalGzL}, we evaluate the order-parameter correlation function, from which we obtain the transverse current-current correlator.
The magnetic noise $\mathcal{N}_{\mathrm{m}}$ generated by a single 2D (bi)layer then follows from Eq.~\eqref{eq:NoiseEq}, and takes the following scaling form above $T_c$: 
\begin{subequations}
    \begin{align}
        \mathcal{N}_{\mathrm{m}}
        &= \frac{\mu_0^2 k_B T}{4\pi} \sigma_{\mathrm{m}}(0)
        \int_0^{\infty} d q\, q\, e^{-2 q z_{\rm NV}}
        F_{\mathrm{m}}\left(q \xi_{\mathrm{m}}\right), \\
        \sigma_{\mathrm{T}}^{\mathrm{m}}(q)
        &= \sigma_{\mathrm{m}}(0) F_{\mathrm{m}}\left(q \xi_{\mathrm{m}}\right),
        \text { where } \sigma_{\mathrm{m}}(0)
        = \frac{1}{8\pi}\left(\frac{e^*}{\hbar}\right)^2
        k_B T \tau_{\mathrm{m}}, \\
        F_{\mathrm{m}}(\tilde{q}=q\xi_{\mathrm{m}})
        &= \frac{2}{\tilde{q}^2}\left[
        \ln \frac{16}{\left(4+\tilde{q}^2\right)^2}
        + \frac{4\tilde{q}}{\sqrt{4+\tilde{q}^2}}
        \tanh^{-1}\left(\frac{\tilde{q}}{\sqrt{4+\tilde{q}^2}}\right)
        \right],
    \end{align}
\end{subequations}
where $\sigma_{\rm m}(0)$ is the uniform ($q = 0$) conductivity, and $F_{\rm m}(q\xi_{\rm m})$ is a scaling function for the nonlocal conductivity $\sigma_{\rm m}(q)$, normalized such that $F_{\rm m}(0) =1$~\cite{cheng2026qubit}.
The (in-plane) correlation length $\xi_{\rm m}$ on the metallic side may be written in terms of the zero-temperature correlation length $\xi_0$ as $\xi_{\mathrm{m}}(T)=\xi_0 \sqrt{2T_c /(T-T_c)}$~\cite{tinkham2004introduction} ($\xi_0 = 1.6$ nm for BSCCO).
Further, it is convenient to write the scattering timescale $\tau_{\mathrm{m}} \propto (T - T_c)^{-1}$ as $\tau_{\mathrm{m}}=C_{\mathrm{m}} \cdot \hbar /\left[k_B\left(T-T_c\right)\right]$, where $C_{\rm m}$ is a dimensionless number to be determined by fitting with the experimental data.

\begin{figure*}[h!]
    \centering
    \includegraphics[width=0.45\textwidth]{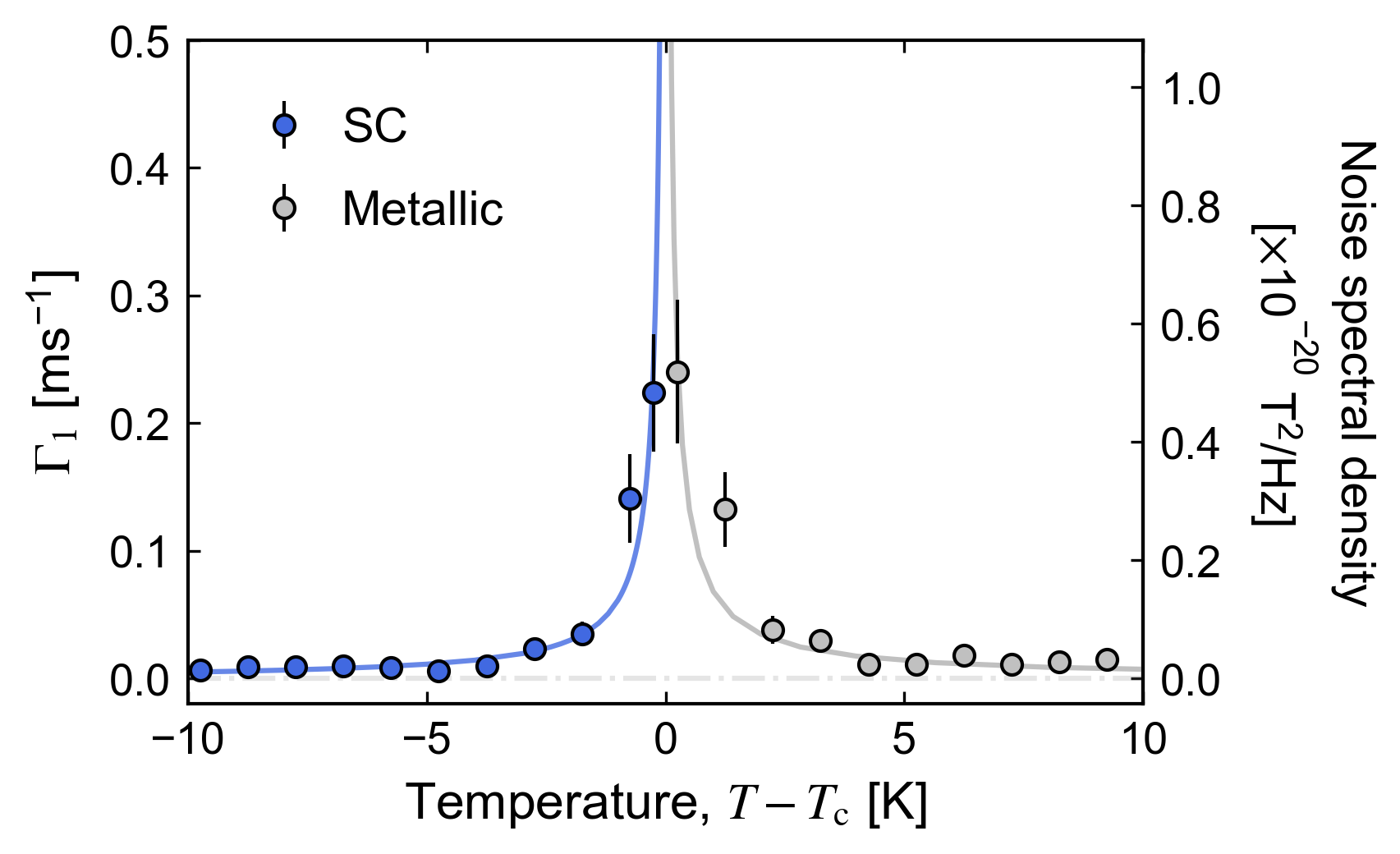}
    \caption{{\bf Theoretical fits to the relaxation rate at zero magnetic field.}
    The measured $\Gamma_1(T)$ exhibits a two-sided divergence near $T = T_c$ due to critical fluctuations, where the theoretical fits based on order-parameter Langevin dynamics are presented as the solid curves.
    The right axis labels the corresponding magnitude of the noise spectral density.
    Error bars represent the 1$\sigma$ uncertainty from fitting.
    }
    \label{fig:TheorySM}
\end{figure*}

Assuming that correlations between different BSCCO bilayers are negligible, we obtain the total relaxation rate by summing the contributions from 120 bilayers.
The nearest bilayer is taken to be $40$ nm from the NV center, and adjacent bilayers are separated by $1.55$~nm as discussed above.
The resulting total magnetic noise is then given by
\begin{equation}
    \mathcal{N}_{\mathrm{m}}^{\mathrm{tot}}
    = \sum_{n=0}^{120}
    \mathcal{N}_{\mathrm{m}}(z_{\rm NV}=40+1.55\,n\mathrm{~nm}) \,.
\end{equation}
The sample-induced $T_1$ thus obtained is plotted in Fig.~3(e) as a function of $T - T_c$ in the main text on a log-log scale, and also in Figure~\ref{fig:TheorySM} on a linear scale. 

As discussed in the main text, the calculated $T_1$ noise scales near the transition as $(T-T_c)^{-1}$ in agreement with experimental observations.
Furthermore, fitting the data in the range $T-T_c\leqslant 10~$K, we obtain $C_{\mathrm{m}} = 0.829$, which 
corresponds to the decay time is $\tau_{\rm m}^{\text{fit}} \approx 2.11 \, \tau_{\text{GL}} = 2.11 \tau_{\rm m}^{\rm wc}$.

\textbf{Superconducting Side}:
On the superconducting side of the transition, the order parameter amplitude has a finite expectation value, given by  $\langle |\psi| \rangle=\sqrt{M_0}$.
It is convenient to represent the order parameter in terms of amplitude and phase fluctuations as
\begin{equation}
    \psi(\bm{\rho},t)=\sqrt{M_{0}+\chi(\bm{\rho},t)}e^{i\theta(\bm{\rho},t)} \,,
    \label{Eq:AmpPhaseRepresentation}
\end{equation}
where $M_0$ represents the homogeneous condensate density, which can be approximated by its mean-field value, and $\chi$ and 
$\theta$ are the density and phase fluctuation fields, respectively.
In particular, phase fluctuations only produce longitudinal currents, and therefore both amplitude and phase fluctuations enter in the evaluation of the transverse current-current correlator \eqref{Eq:TransverseConductivity}.

Substituting \eqref{Eq:AmpPhaseRepresentation} into the time-dependent Ginzburg--Landau equation yields coupled Langevin dynamics for the amplitude and phase fields, where the two fields decouple at leading order.
We find that the relaxation time of the amplitude mode is given by $\tau_{\rm sc} = 1/(2\gamma |r|) = 1/(2\gamma uM_0)$, while its correlation length is $\xi_{\rm sc}(T) = \sqrt{K/(2 |r|)}$, implying that $\nu = 1/2$ and $z = 2$ on the superconducting side as well. 
Using the equations of motion to compute the transverse current-current correlations, we obtain the following scaling form for the magnetic noise $\mathcal{N}_{\mathrm{sc}}$ generated by a 2D superconducting (bi)layer below $T_c$~\cite{cheng2026qubit}:
\begin{subequations}
    \begin{align}
        \mathcal{N}_{\mathrm{sc}}
        &= \frac{\mu_0^2 k_B T}{4\pi} \sigma_{\mathrm{sc}}(0)
        \int_0^{\infty} d q\, q\, e^{-2 q z_{\rm NV}}
        F_{\mathrm{sc}}\left(q \xi_{\mathrm{sc}}\right), \\
        \sigma_{\mathrm{T}}^{\mathrm{sc}}(q)
        &= \sigma_{\mathrm{sc}}(0) F_{\mathrm{sc}}\left(q \xi_{\mathrm{sc}}\right),
        \text { where } \sigma_{\mathrm{sc}}(0)
        = \frac{\ln 2}{\pi}\left(\frac{e^*}{\hbar}\right)^2
        k_B T \tau_{\mathrm{sc}}, \\
        F_{\mathrm{sc}}(\tilde{q}=q\xi_{\mathrm{sc}})
        &= \frac{1}{\tilde{q}^2\ln 2}\left[
        \ln\left(4\frac{1+\tilde{q}^2}{2+\tilde{q}^2}\right)
        - \frac{\ln\left(2+\tilde{q}^2\right)}{1+\tilde{q}^2}
        \right],
    \end{align}
\end{subequations}
where $\sigma_{\rm sc}(0)$ is the zero-momentum paramagnetic conductivity, and $F_{\mathrm{sc}}(q \xi_{\rm sc})$ is a scaling function for the nonlocal conductivity on the superconducting side.

Analogous to our noise computation on the metallic side, we write the scattering time on the superconducting side as  $\tau_{\mathrm{sc}}=C_{\mathrm{sc}} \cdot \hbar /\left[k_B\left(T_c-T\right)\right]$,  and note that $\xi_{\mathrm{sc}}(T)=\xi_0 \sqrt{T_c /(T_c-T)}$ with $\xi_0=1.6~$nm for BSCCO. 
Subsequently, we find the total magnetic noise $\mathcal{N}_{\mathrm{sc}}^{\mathrm{tot}}$ on the superconducting side by summing over each CuO$_2$ bilayer in the BSCCO sample:
\begin{equation}
    \mathcal{N}_{\mathrm{sc}}^{\mathrm{tot}}
    = \sum_{n=0}^{120}
    \mathcal{N}_{\mathrm{sc}}(z_{\rm NV}=40+1.55\,n\mathrm{~nm}) \,.
\end{equation}
Using Eq.~\eqref{Eq:DepolarizationRate}, we obtain the depolarization rate from the transverse magnetic noise.
The result is displayed in Fig.~3(e) in the main text as a function of $|T-T_c|$ on a log-log scale, and also in Figure~\ref{fig:TheorySM}  as a function of $T-T_c$.
Fitting the data in the range $|T-T_c| \leqslant 10~$K, we find $C_{\rm sc}  = 0.124$, corresponding to $\tau_{\rm sc}^{\text{fit}}/\tau_{\rm sc}^{\text{wc}}=0.63$.

\methodsubsection{Finite-field noise from vortex dynamics}
Next, we consider the magnetic noise at a non-zero external field $H$ arising from the motion of field-induced vortices, which becomes the predominant noise mechanism in a broad temperature window below $T_c$.
For a qubit with quantization axis $\hat{\z}$, depolarization is caused by in-plane magnetic field fluctuations, captured by the rotationally invariant noise tensor component $\mathcal{N}_{xx}=\mathcal{N}_{y y}=\mathcal{N}_{\rm T}/2$~\cite{cheng2026qubit}.
For a sample with $N$ mobile vortices, we may compute the net magnetic field fluctuation as $\delta \B(t) = \sum_{i=1}^{N} \delta \B_{i}(t)$, where $\delta \B_{i}(t)$ denotes the fluctuating field at the NV due to the motion of vortex $i$. Accordingly, we can write
\begin{equation}
    \mathcal{N}_{xx}(\Omega)
    = \sum_{i,j=1}^N\int_{-\infty}^{\infty} d t\,e^{i\Omega t}
    \left\langle \delta B_{i,x}(t) \cdot \delta B_{j,x}(0) \right\rangle_T \,,
    \text{ where } \delta B_{i,x}(t)
    = \delta \B_{i}(t) \cdot \hat{\x}.
    \label{eq:noise_B_x}
\end{equation}
Let the NV coordinate be  $(\bm{\rho}_0,z_0)$ and the center of the vortex $i$ at time $t$ be $(\bm{R}_i+\u_i(t),0)$, $\bm{R}_i$ being its thermal equilibrium position and $\langle\u_i(t)\rangle_T=0$.
We define the equilibrium in-plane separation between the NV and vortex $i$ as $\bm{\rho}_i\equiv\bm{\rho}_0-\bm{R}_i$. 
We may Taylor expand the magnetic field from vortex $i$ about the equilibrium separation $\bm{\rho}_i$ to find the fluctuating component of the field $\delta \B_i(t)$ due to vortex motion.
Denoting the equilibrium field as $\bm{B}_{0}(\bm{\rho}_i,z)$, we find:
\begin{equation}
    B_{i,x}(\bm{\rho}_i - \u_i(t),z) =  B_{0,x}(\bm{\rho}_i,z) + \bm{g}_i \cdot \u_i(t) +
    \mathcal{O}(u_i^2) ~,
\end{equation}
where $\bm{g}_i = - \bm{\nabla}_{\parallel} B_{0,x} \left( \bm{\rho},z \right) \big|_{\bm{\rho}=\bm{\rho}_i}$, and $\bm{\nabla}_{\parallel}=\hat{\bm\rho} \partial_\rho+\hat{\bm\theta}\rho^{-1} \partial_\theta$ denotes the transverse (in-plane) gradient of the field components.

\begin{samepage}
Assuming identical statistical behavior and uncorrelated fluctuations among distinct vortices, the transverse noise reduces to a sum over single-vortex displacement self-correlators, which can be further simplified using in-plane rotational invariance of these correlators~\cite{cheng2026qubit}:
\begin{equation}
    \mathcal{N}_{x x}(\Omega)
    = \sum_{i=1}^N\left|\bm{g}_i\right|^2
    \frac{\langle \u_{\Omega} \cdot \u_{-\Omega}\rangle_T}{2},
    \text{ where } \langle \u_{\Omega} \cdot \u_{-\Omega}\rangle_T
    = \int_{-\infty}^{\infty} d t\,e^{i\Omega t}
    \left\langle \u_t \cdot \u_0 \right\rangle_T \,.
\end{equation}
\end{samepage}

The average noise $\langle\langle\mathcal{N}_{x x}(\Omega)\rangle\rangle$ is evaluated using the configuration average over spatial arrangements of vortices:
\begin{equation}
    \begin{aligned}
        \langle\langle\mathcal{N}_{x x}(\Omega)\rangle\rangle
        &= \frac{\langle \u_{\Omega}\cdot\u_{-\Omega}\rangle_T}{2}
        \langle\langle\sum_{i=1}^N |\bm{g}(\bm{R}_i)|^2\rangle\rangle \\
        &= \frac{\langle \u_{\Omega}\cdot\u_{-\Omega}\rangle_T}{2}
        \int d^2 R \,
        \langle\langle\sum_{i=1}^N
        \delta^{(2)}\left(\bm{R}-\bm{R}_i\right)\rangle\rangle
        \left|\bm{g}(\bm{R})\right|^2 \\
        &\simeq \frac{\langle \u_{\Omega}\cdot\u_{-\Omega}\rangle_T}{2}
        n_v\int_{\text{sample}}\left|\bm{g}(\bm{R})\right|^2 \,.
    \end{aligned}
\end{equation}
Here, $\langle\langle\sum_{i=1}^N \delta^{(2)}\left(\bm{R}-\bm{R}_i\right)\rangle\rangle$ stands for the mean vortex density at position $\R$ in the sample plane. For a macroscopically uniform distribution, this quantity reduces to the global average vortex density $n_v = H/\Phi_0$. 
This $H$-linear scaling of the noise spectral density implies, via Eq.~\eqref{Eq:DepolarizationRate}, that the  corresponding depolarization rate $\Gamma_1 =3 (g \mu_B/\hbar)^2 \langle\langle\mathcal{N}_{xx}\rangle\rangle \propto n_v \propto H$, i.e., $\Gamma_1(H)$ scales linearly with the external magnetic field $H$.

\bookmarksetupnext{level=subsection}
\putbib[references]

\end{bibunit}


\clearpage
\onecolumngrid

\bookmarksetup{startatroot}
\bookmarksetupnext{level=section}
\majorsection{Supplementary Information}{supp}

\suppnumbering

\supplementcontents

\setbiblinksuffix{-supp}
\begin{bibunit}[apsrev4-2-etal]

\suppsection{Nitrogen-vacancy Center Basics}

The nitrogen-vacancy (NV) center in diamond is a color defect in diamond crystals, consisting of a nitrogen atom replacing a carbon atom adjacent to a vacancy in the carbon lattice~\cite{doherty2013nitrogen}.
The electronic ground state of a negatively charged NV center forms a spin-1 triplet state.
In the absence of external disturbances, the $|m_s=0\rangle$ sublevel is separated from the degenerate $|m_s=\pm1\rangle$ sublevels by a zero-field splitting $D_\mathrm{gs} = 2.87$~GHz, along the quantization axis $\hat{z}$ in the N-V direction.
In the presence of a magnetic field, the axial component $B_z$ lifts the degeneracy between $|m_s=\pm1\rangle$ states with a Zeeman splitting $2\gamma_e B_z$, where $\gamma_e=2.8$~MHz/G is the electron spin gyromagnetic ratio.
Additionally, the local electric charges around NV centers will couple to the system~\cite{mittiga2018imaging}, represented by the Hamiltonian:
\begin{equation}
    \mathcal{H}_E = \Pi_z S_z^2 + \Pi_x (S_y^2-S_x^2) + \Pi_y (S_x S_y + S_y S_x),
\end{equation}
where $\{S_x, S_y, S_z\}$ are the spin-1 operators, $\Pi_{\{x,y\}}=d_\perp E_{\{x,y\}}$ and $\Pi_z=d_\parallel E_z$ characterize the coupling strength to the electric field $\bm{E}$ with susceptibilities $\{d_\perp, d_\parallel\}$ = \{17, 0.35\}~Hz$~\cdot~$cm/V~\cite{vanoort1990electric}.
Given that $d_\perp \gg d_\parallel$, the axial term $\Pi_z$ can be neglected.
Meanwhile, the transverse terms only couple to the $|m_s=\pm1\rangle$ states, producing new eigenstates:
\begin{align}
    |+\rangle &= \frac{1}{\sqrt{2}}\left(|m_s=+1\rangle + e^{i\theta}|m_s=-1\rangle\right) \\
    |-\rangle &= \frac{1}{\sqrt{2}}\left(|m_s=+1\rangle - e^{i\theta}|m_s=-1\rangle\right)
\end{align}
where $\theta = \arctan{(\Pi_y/\Pi_x)}$ represents the transverse orientation of electric field.
This interaction introduces an energy splitting of $2\Pi_\perp = 2\sqrt{\Pi_x^2+\Pi_y^2}$.
As a result, the local-charge-induced splitting combines in quadrature with the Zeeman splitting, leading to a total splitting of $\Delta = \sqrt{(2\Pi_\perp)^2+(2\gamma_e B_z)^2}$ shown in the ODMR spectrum.

\suppsubsection{Optically detected magnetic resonance}
In this work, we perform continuous-wave optically detected magnetic resonance (ODMR) spectroscopy to probe the spin transitions of the NV centers.
During the measurement, both the laser excitation and microwave driving fields are applied continuously, while the frequency of the microwave field is swept.
The NV spins are polarized into their initial state $|m_s=0\rangle$ by the optical pumping.
When the microwave is off-resonant, the spin population remains predominantly in $|m_s=0\rangle$.
As the microwave frequency is swept on resonance with one of the spin transitions, the spin population is driven into the $|m_s=\pm1\rangle$ states, resulting in a decrease of NV fluorescence.

Due to the tetrahedral symmetry of the diamond lattice, the projection of the external magnetic field along the NV axis, $B_z$, depends on its specific crystallographic orientation.
Therefore, the NV ensemble can be divided into four subgroups.
Throughout our experiment, the external magnetic field is applied normal to the surface of the [111]-cut diamond, aligning with one NV group.
In this configuration, the ODMR spectrum exhibits four distinct resonant peaks.
The NV subgroup aligned with the external field experiences the largest magnetic field projection, giving rise to the two outer resonances separated by a splitting $\Delta$.
Meanwhile, the NV groups in the other three directions share an identical projection and remain degenerate, contributing to the inner peaks.

\begin{figure*}[t!]
    \centering
    \includegraphics[width=1\textwidth]{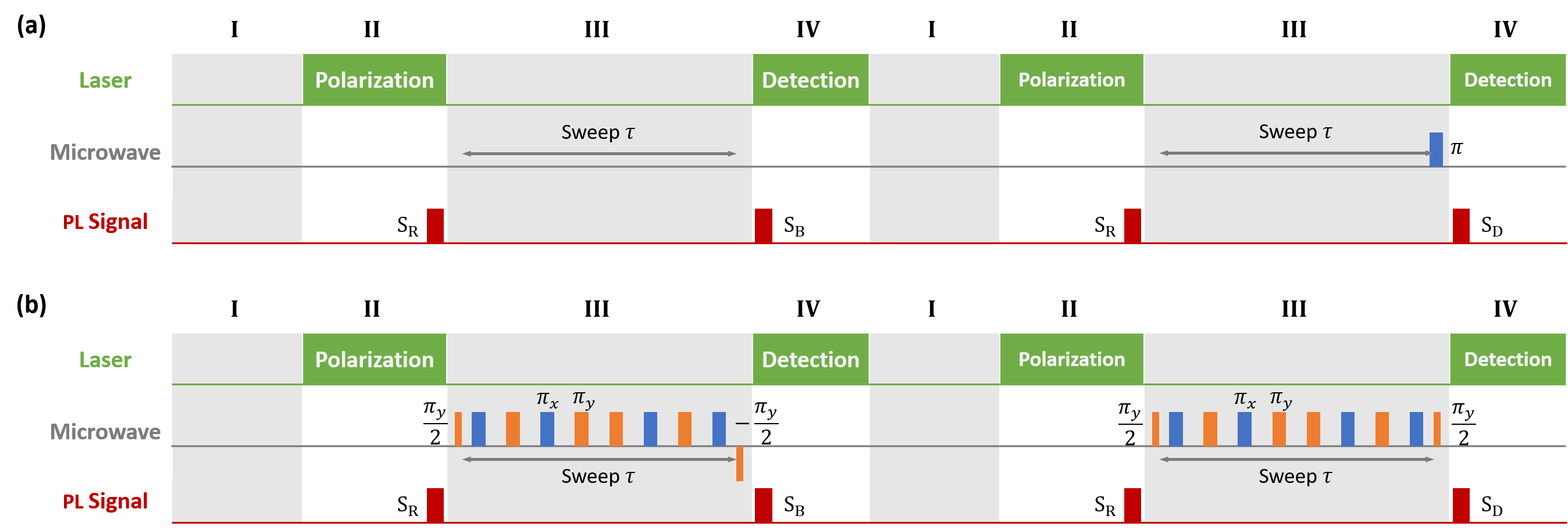}
    \caption{{\bf Pulse sequences of differential measurements.}
    Schematic of (a) $T_1$ and (b) XY-8 pulse sequences used in the experiment.
    The $\pi$-pulse at the end of the microwave sequence is applied to swap the population from the dark state $|m_s=\pm1\rangle$ to the bright state $|m_s=0\rangle$.
    }
    \label{fig:fig_seqs}
\end{figure*}

\suppsubsection{Spin relaxation}
To characterize the spin relaxation of the NV ensemble, we perform a differential sequence to measure the lifetime of $|m_s=0\rangle$ state, while mitigating the NV charge dynamics effect under laser pumping~\cite{choi2017depolarization}, see Figure~\ref{fig:fig_seqs}(a).
To be specific, we first let NV charge dynamics reach equilibration by waiting for 100~$\mu$s without laser illumination (I).
Next, a 15~$\mu$s laser pulse is applied to initialize the NV centers to $|m_s=0\rangle$ state (II).
A reference readout $\mathrm{S_R}$ is taken at the end of the pulse.
After the polarization, we pause for a variable time period $\tau$ for spin relaxation (III), then turn on the laser to detect the spin population of the bright state $|m_s=0\rangle$ as $\mathrm{S_B(\tau)}$ (IV).
Finally, we repeat the I-IV procedures but with a $\pi$-pulse at the end of time $\tau$ to swap the population from the dark state $|m_s=\pm1\rangle$, recorded as $\mathrm{S_D(\tau)}$, to $|m_s=0\rangle$.
As a result, the relaxation time $T_1$ is extracted by fitting the fluorescence contrast to a stretched exponential decay
\begin{equation}
    C(\tau) = \frac{\mathrm{S_B(\tau)-\mathrm{S_D(\tau)}}}{\mathrm{S_R}} \propto
    e^{-\left(\frac{\tau}{T_1}\right)^{\alpha_1}} ,
\end{equation}
where the stretched power $\alpha_1 = 1/2$ according to a pioneering study~\cite{choi2017depolarization}.

To quantitatively analyze the relaxation rate induced by the fluctuation noise, we treat the overall decay as a combination of an intrinsic decay and the noise-related decay,
\begin{equation}
    C(\tau) \sim e^{-\sqrt{\tau/T_1}}
    = e^{-\sqrt{\Gamma_1^\mathrm{total}~\tau}}
    = e^{-\sqrt{\Gamma_1^\mathrm{int}~\tau}} \cdot e^{-\sqrt{\Gamma_1^\mathrm{noise}~\tau}}.
\end{equation}
Then the decay rate $\Gamma_1^\mathrm{noise}$, shorten as $\Gamma_1$ in the main text, can be written as
\begin{equation}
    \Gamma_1 = \left(\sqrt{\Gamma_1^\mathrm{total}}-\sqrt{\Gamma_1^\mathrm{int}}\right)^2,
\end{equation}
with $\Gamma_1^\mathrm{total}$, $\Gamma_1^\mathrm{int}$ respectively extracted from the measurements at NV centers under and away from the BSCCO flake.

For $T_1$-based noise spectroscopy, the spin depolarization dynamics are determined by the magnetic noise spectrum at the NV resonance frequencies $\omega\simeq 2\pi \times (D_\mathrm{gs}\pm\Delta/2)$, leading to a sharp frequency filter function $W_\tau(\omega)$ peaked in the GHz range.

\suppsubsection{Spin decoherence and dynamical decoupling}
We apply a dynamical decoupling pulse sequence, XY-8~\cite{gullion1990new}, to probe the spin decoherence properties of the NV ensemble, see Figure~\ref{fig:fig_seqs}(b).
Compared to the commonly used Spin Echo pulse sequence, instead of a single refocusing $\pi$-pulse in the middle, XY-8 employs a series of $\pi$-pulses with alternating phases.
As a result, XY-8 further suppresses the local spin noise and is self-compensating for pulse errors.

Similar to spin relaxation, the spin coherence time $T_2$ follows a stretched exponential decay profile
\begin{equation}
    C(\tau) \propto e^{-\left(\frac{\tau}{T_2}\right)^{\alpha_2}},
\end{equation}
with a stretched power $\alpha_2 = 3/2$ for NV ensemble~\cite{bauch2020decoherence}.

The decoherence rate induced from the fluctuation noise, $\Gamma_2$, takes the form
\begin{equation}
    \Gamma_2 = \left((\Gamma_2^\mathrm{total})^{\frac{3}{2}} -
    (\Gamma_2^\mathrm{int})^{\frac{3}{2}}\right)^{\frac{2}{3}},
\end{equation}
where $\Gamma_2^\mathrm{total}$, $\Gamma_2^\mathrm{int}$ are extracted from NV centers under and away from the BSCCO flake.
To ensure the non-negative value inside the parentheses, any negative result is capped at zero.

In contrast to $T_1$-based noise spectroscopy, the filter function of $T_2$-based noise spectroscopy can be tailored by the time interval $\tau$, see supplementary material in Ref.~\cite{machado2023quantum}:
\begin{equation}
    W_{\tau}^\mathrm{(XY8)}(\omega) \sim \frac{\sin^2{\frac{\omega\tau}{2}}}{\omega^2} \cdot
    \frac{\sin^4{\frac{\omega\tau}{32}}}{\cos^2{\frac{\omega\tau}{16}}}.
\end{equation}
It peaks at $\omega \simeq 8\pi/\tau$, with width $\sim \pi/\tau$.
In this work, the time $\tau$ is swept up to $\sim 10~\mu$s, enabling a noise spectroscopy that is sensitive to the MHz range.

\suppsection{Experimental Setup and Sample Preparation}

\suppsubsection{Setup apparatus}
We perform NV measurements in a commercial low-vibration cryostat (Fournine Design SK200) which has a base temperature below 4~K.
A closed-cycle helium compressor (Sumitomo F-40L) as well as a cryogenic temperature controller (LakeShore Model 336) and temperature sensors with integrated heaters are employed to maintain the sample at the desired temperature in the range from 4~K to 350~K.

To apply an external magnetic field along the out-of-plane direction of the BSCCO flake, a Neodymium-45 permanent magnet (SuperMagnetMan Cyl0505) is mounted on an XYZ translation stage (Newport 460A-XYZ) outside of the cryostat, oriented perpendicular to the diamond surface.
The magnetic field also lift the degeneracy between the NV $|m_s = \pm1\rangle$ states, forming effective two-level systems $|m_s = 0,-1\rangle$ or $|m_s = 0,+1\rangle$ for further spin controls.
In the small-field circumstance ($H<10$~G), the permanent magnet is replaced by a solenoid magnet (TOMSHIELE TS-P30/22), driven by a DC power supply (B\&K Precision 9131B) to precisely control the field strength.

The cryostat is equipped with optical access windows and integrated with a home-built confocal scanning microscope as well as a widefield imaging microscope.
In the confocal setup, the initialization and detection of NV center spins are driven by laser light with wavelength $\lambda = 532$~nm (Millennia eV CW-DPSS-5W).
The laser is modulated by an acousto-optic modulator (AOM, G$\&$H AOMO 3110-120) in a double-pass configuration to achieve an on/off ratio exceeding $10^5:1$.
The lateral scanning is performed using an X-Y galvanometer (Thorlabs GVS212), in combination with a 4$f$-telescope setup.
A piezo objective scanner (Physik Instrumente P-721.SL2) controls the vertical position with a 100~$\mu$m travel range.
The laser beam is focused onto the sample by an objective lens (Mitutoyo Plan Apo 20x/0.42 NA) mounted on the piezo scanner, achieving a diffraction-limited spot size with a diameter less than $1~\mu$m.
The NV fluorescence is collected by the same objective, separated from the laser beam via a dichroic mirror (Semrock FF560-Di01) and filtered through a long-pass filter (Semrock BLP01-633R) before being detected by a single-photon counting module (Excelitas SPCM-AQRH-63-FC).
A data acquisition device (National Instruments USB-6343) is used for fluorescence counting and subsequent data processing.
In the widefield setup, a separate laser source with built-in modulation (Cobolt 06-MLD 520~nm) is used to provide the excitation beam with wavelength $\lambda = 520$~nm.
The laser is expanded by a beam expander (Edmund Optics Scorpii-6X) and then focused onto the objective lens to illuminate the diamond.
The NV fluorescence image is collected by the objective, filtered through a dichroic mirror and a long-pass filter, and finally mapped to a CMOS camera (iDS UI-3240CP-NIR) via an infinity-corrected tube lens (Thorlabs TTL200-B).
A mirror on a flip mount is used to switch between two setups.

\begin{figure*}[h!]
    \centering
    \includegraphics[width=0.8\textwidth]{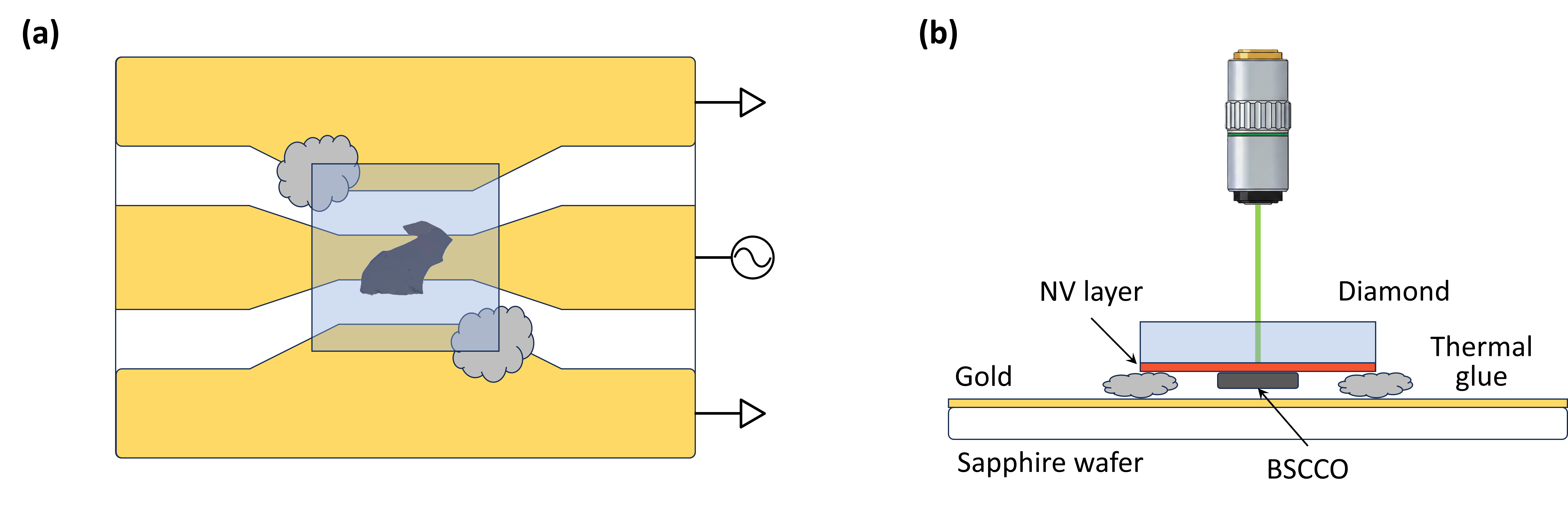}
    \caption{{\bf Microwave delivery.}
    (a) Schematic of the coplanar waveguide (CPW) in top view.
    (b) Schematic of the sample mounting with CPW in side view.
    Both schematics are not to scale.
    }
    \label{fig:fig_CPW}
\end{figure*}

The microwave driving field is generated from a signal generator (Stanford Research Systems SG384), shuttered by a switch (Mini-Circuits ZASWA-2-50DRA+) to prevent potential leakage, and amplified by a high-power amplifier (Mini-Circuits ZHL-15W-422-S+).
A custom-designed coplanar waveguide (CPW), mounted on the cold finger of the cryostat, is used to deliver the microwave to the NV ensemble, see Fig.~\ref{fig:fig_CPW}.
The CPW is fabricated with a 160~nm-thick Au pattern and a 10~nm Cr adhesion layer deposited on a 400~$\mu$m-thick sapphire wafer, and is designed for a matched impedance of 50~$\Omega$.
The transmitted microwave is then delivered outside the cryostat and terminated by a 40~dB attenuator (Mini-Circuits BW-S40W20+) with a 50~$\Omega$ terminator (Mini-Circuits ANNE-50L+) at the end.
All the equipment is synchronized via a programmable multi-channel pulse generator (SpinCore PulseBlasterESR-PRO 500) with a 2~ns temporal resolution.

\suppsubsection{Thermal analysis of laser and microwave heating}
Our experiments involve applying optical excitation and microwave driving fields in close proximity to a superconducting sample.
In this section, we address potential heating concerns and estimate the thermal load on the system.

First of all, the cooling power of our cryostat is approximately 200~mW at the base temperature, which provides a substantial thermal budget.
Crucially, the BSCCO flake is exfoliated directly onto the diamond surface, which ensures efficient thermal contact between the sample and the cooling system.
In contrast, for pulsed confocal measurements, the input laser power is maintained below 0.5~mW.
For the microwave input, as described in the previous section, our custom-designed, impedance-matched coplanar waveguide ensures that the majority of the microwave power is transmitted through the cryostat and dissipated at room temperature, rather than within the cryogenic environment.

Furthermore, the duty cycle of both the laser and microwave pulses during $T_1$ and $T_2$ measurements is extremely low, effectively minimizing localized heating.
At the beginning of each measurement sequence, there is a 100~$\mu$s waiting period for both signal-bright and signal-dark, as depicted in Fig.~\ref{fig:fig_seqs}.
This period serves two purposes: it allows the NV centers to reset to their charge equilibrium state and significantly reduces the duty cycle, thus the time-averaged power from both the laser and microwave.
For instance, the $\pi$-pulse duration in the measurements is around 50~ns or less, while the minimum total length of the most microwave-intensive sequence, XY8, is more than 250~$\mu$s for a differential measurement.
Therefore, the maximum duty cycle for a XY8 sequence should be less than $[50~{\rm ns} \times (8+1) \times 2] ~/~ (250~{\rm \mu s}) \approx 0.4\%$.
The microwave duty cycle is even orders of magnitude lower in $T_1$ sequence, where only a single $\pi$-pulse is applied within a sequence that can extend to tens of milliseconds.

For the continuous-wave ODMR, we apply a much lower laser power ($\sim 0.1$~mW, 5 times less than the pulsed measurement) and microwave power ($\sim 5$~mW, about 27~dB less than the pulsed measurement) in the experiments.
We note the resonant frequencies used in $T_1$ and $T_2$ sequences are determined from a pulsed-ODMR measurement, where the duty cycle is controlled to be less than 10\% (laser, effective time-averaged power $\sim 0.01$~mW) and 0.03\% (microwave, effective time-averaged power $\sim 1.5~\mu$W).

In the case of a widefield experiment, the total input laser power is 30~mW, but it is distributed over a large field of view of $\sim 200 \times 150~\mu$m.
This results in a laser intensity of $\sim 1~{\rm \mu W / \mu m^2}$, which is significantly lower than the intensity at the diffraction-limited spot in the confocal setup.
The microwave power remains comparable to that used in confocal measurements.
Based on these estimates, we believe that the induced heating from both laser and microwave sources is negligible compared to the cooling capacity of the cryostat.

\begin{figure*}[h!]
    \centering
    \includegraphics[width=0.9\textwidth]{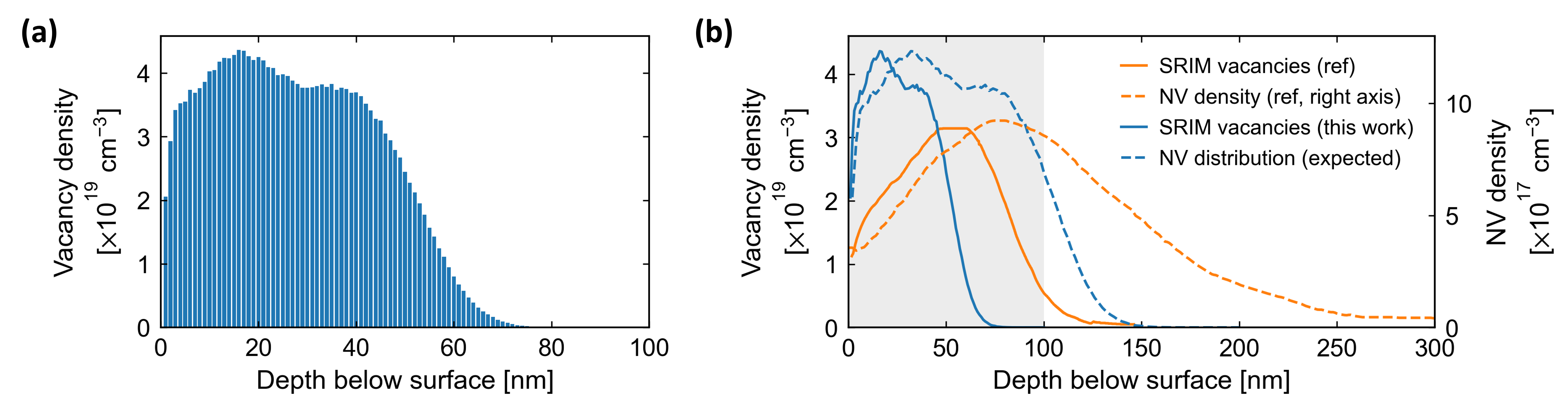}
    \caption{{\bf The vacancy depth distribution.}
    (a) The vacancy density as a function of the depth below the diamond surface, simulated using Stopping and Range of Ions in Matter (SRIM).
    The implanted vacancies are concentrated in the first $\sim 50$~nm layer beneath the diamond surface.
    (b) Based on Ref.~\cite{santori2009vertical}, we expect that after diffusion during annealing, most NV centers are concentrated within $\lesssim 100$~nm from the surface.
    }
    \label{fig:fig_SRIM}
\end{figure*}

\suppsubsection{Creation of NV centers}
In this work, we use a double-side polished single crystal type-Ib diamond plate in the shape of 2.5~mm $\times$ 2.5~mm $\times$ 0.3~mm with [111]-cut surfaces (Sumitomo).
The diamond is synthesized through a high-pressure high-temperature (HPHT) process, with $[\mathrm{N}] < 200$~ppm~\cite{zuo2021synthetic}.
We employ $^{12}\mathrm{C}^+$ ion implantation (CuttingEdge Ions, LLC) at 12~keV energy with a dosage of 2.25~$\times 10^{11}$~cm$^{-2}$ and 30~keV energy with a dosage of 6.75~$\times 10^{11}$~cm$^{-2}$ sequentially.
We perform a detailed calculation with full damage cascades using Stopping and Range of Ions in Matter (SRIM)~\cite{ziegler2010srim} to obtain the total created NV defect density distribution.
As a result, a $\sim 50$~nm layer of vacancies is generated below the plate surface, see Fig.~\ref{fig:fig_SRIM}(a).
After implantation, the diamond is placed in vaccum ($< 10^{-6}$~mbar) and annealed in a home-built furnace with a recipe same to previous work~\cite{hsieh2019imaging}: 12~hours ramp to 400$^\circ\mathrm{C}$, dwell for 8~hours, 12~hours ramp to 800$^\circ\mathrm{C}$, dwell for 8~hours, 12~hours ramp to 1050$^\circ\mathrm{C}$, dwell for 2~hours.
The NV concentration after the annealing process is estimated to be around 1~ppm.

We note that, although the vacancies could diffuse into the diamond during annealing, the high nitrogen density in the type-Ib diamond would limit this process, leading to an NV depth distribution comparable to the initial implantation profile.
For example in Ref.~\cite{santori2009vertical} [Fig.~\ref{fig:fig_SRIM}(b)], 
SRIM simulations indicate that the implanted vacancies are spread over a depth of 100~nm (solid orange curve), while the experimentally extracted NV distribution after annealing shows a longer tail, with the majority of the NV centers remaining within a depth of $\lesssim 200$~nm (dashed orange curve).
Therefore, in our sample with shallower implanted vacancies of $\sim 50$~nm (solid blue curve), we anticipate that the resulting NV centers are well-confined within a depth of 200~nm, and mainly concentrated in the first $\lesssim 100$~nm from the diamond surface (dashed blue curve).

\suppsubsection{BSCCO sample preparation}
The Bi$_2$Sr$_2$CaCu$_2$O$_{8+\delta}$(BSCCO, Bi-2212) single crystals used in this work are grown by the traveling floating zone technique~\cite{wen2008large}.
We exfoliate BSCCO flakes using Scotch tape and transfer them directly onto the diamond surface.
Due to the air-sensitivity of BSCCO, we immediately enclose the sample in a vacuum environment within the cryostat.
To estimate the thickness of the target flake, we exfoliate additional flakes separately and identify several with similar optical contrast to the target.
We then measure their thickness using atomic force microscopy.
The measured thicknesses range from $188\pm5$~nm to $300\pm10$~nm, leading us to estimate that the target flake is approximately $200$~nm thick.
The total air exposure time is kept under an hour, and given the flake’s thickness, this brief exposure is not expected to significantly impact its superconducting properties.

\suppsection{Data Analysis}

\suppsubsection{Stretched exponential analysis of decay rates}
During the extraction of relaxation time ($T_1$) and decoherence time ($T_2$), we have used fixed stretched exponential exponents for the fits to ensure robust and consistent results across all temperatures.
This approach is supported by prior theory and experiment on NV centers, and is also consistent with our experimental data.
For the relaxation time $T_1$ of a near-surface NV ensemble, a stretching exponent of $\alpha_1 = 1/2$ can arise from the averaging over NV centers with a distribution of $T_1$ values~\cite{choi2017depolarization}.
Similarly, for the coherence time $T_2$, an exponent of $\alpha_2 = 3/2$ is expected due to an averaging effect in the presence of a fluctuating spin bath~\cite{bauch2020decoherence}.

Our experimental data are in good agreement with these prior works.
To see this more clearly, we express the $T_1$ and $T_2$ decay curves in the formula:
\begin{equation}
    C(\tau) = A \cdot e^{-(\frac{\tau}{\tau_c})^\alpha},
\end{equation}
where $\tau_c = T_1$ or $T_2$ and $\alpha$ is the stretching exponent.
Taking the logarithm of both sides of the formula gives
\begin{equation}
    \ln{C(\tau)} = \ln{A} - \left(\frac{\tau}{\tau_c}\right)^\alpha .
\end{equation}
Consequently, plotting $-\ln{C(\tau)}$ against $\tau$ on a log-log scale yields a linear relationship where the slope corresponds directly to the exponent $\alpha$.

\begin{figure*}[h!]
    \centering
    \includegraphics[width=\textwidth]{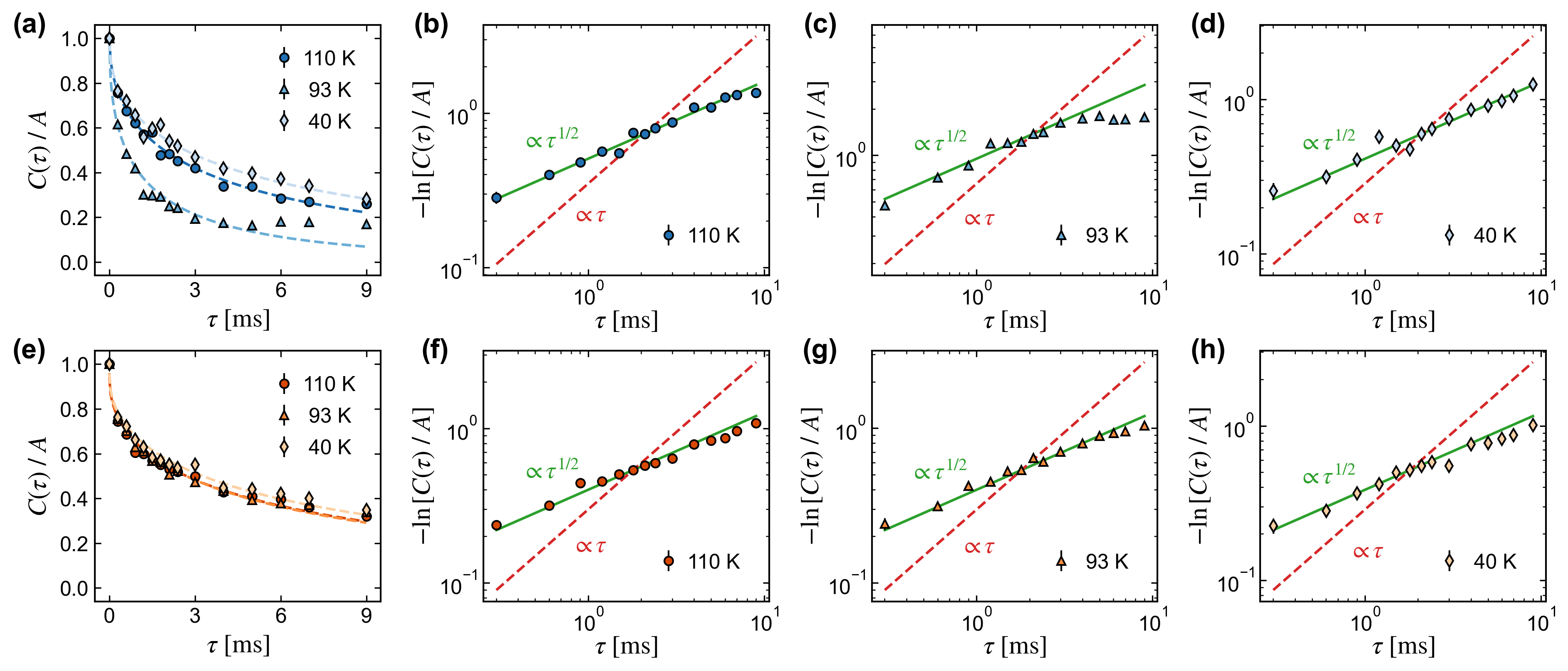}
    \caption{{\bf Stretched exponential analysis of NV relaxation time $T_1$.}
    (a, e) Normalized relaxation curves ($T_1$ decay) for NV ensembles located (a) under the BSCCO flake and (e) far from the sample.
    These data correspond to the measurements shown in Fig.~3(a) in the main text.
    (b-d, f-h) The corresponding decay data are plotted on a log-log scale to determine the stretching exponent $\alpha_1$.
    The data at representative temperatures -- above $T_c$ (110~K), near $T_c$ (93~K), and below $T_c$ (40~K) -- are shown for both sample locations.
    The green lines, representing a slope of 1/2, are a guide to the eye and demonstrate that the decay is consistent with an exponent of $\alpha_1 = 0.5$.
    The dashed red lines show a simple exponential decay (slope of 1) for comparison.
    Error bars represent the photon shot noise, which are barely visible due to statistical averaging.
    }
    \label{fig:fig_stretched_T1}
\end{figure*}

\begin{figure*}[h!]
    \centering
    \includegraphics[width=\textwidth]{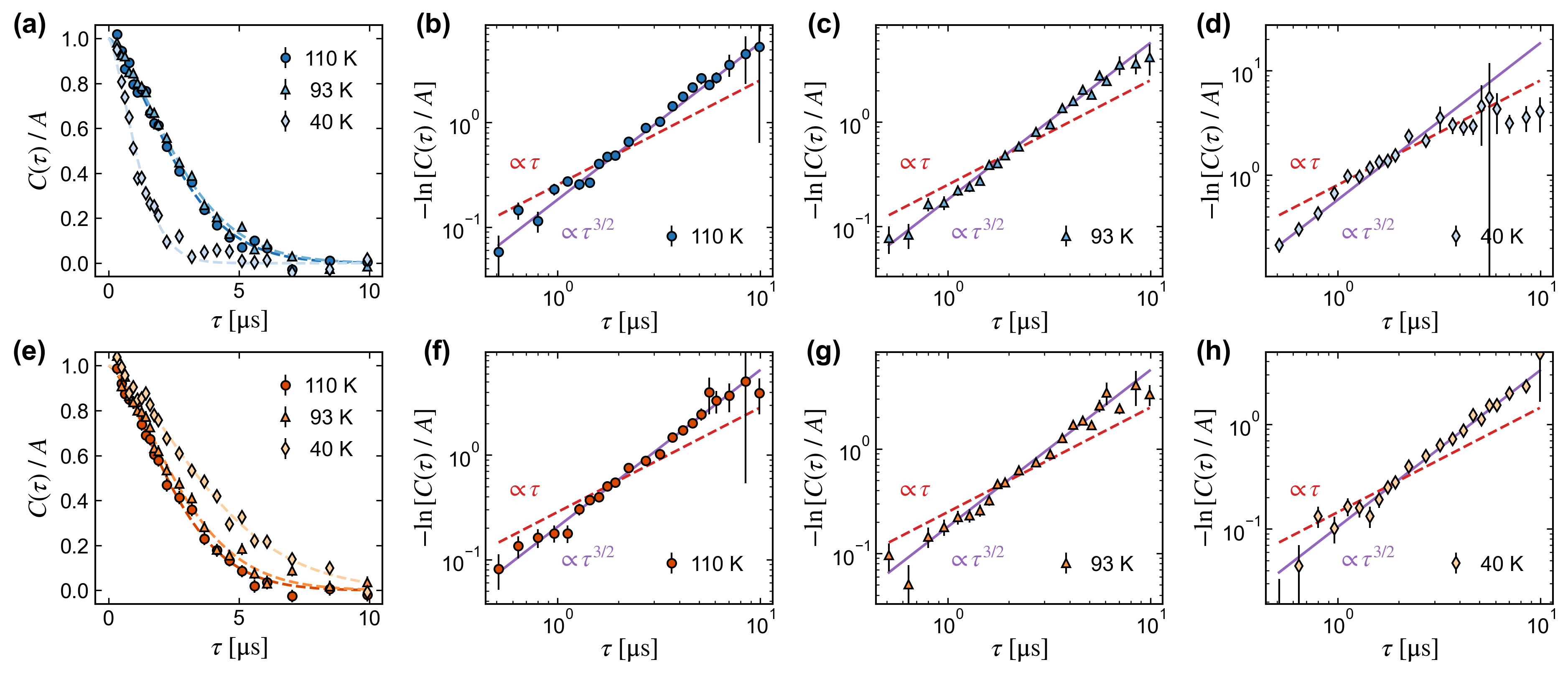}
    \caption{{\bf Stretched exponential analysis of NV decoherence time $T_2^{\rm XY8}$.}
    (a, e) Normalized decoherence curves ($T_2$ decay) for NV ensembles located (a) under the BSCCO sample and (e) far from the sample, measured under a $H = 200$~G bias field.
    (b-d, f-h) The corresponding decay data are plotted on a log-log scale to determine the stretching exponent $\alpha_2$.
    The data are shown at the same representative temperatures as in Fig.~\ref{fig:fig_stretched_T1}.
    The purple lines indicate a 3/2 slope as a guide to the eye, confirming the decay is consistent with an exponent of $\alpha_2 = 1.5$.
    The dashed red lines with a slope of 1 are shown for comparison.
    Error bars represent the photon shot noise, which are barely visible due to statistical averaging.
    }
    \label{fig:fig_stretched_T2}
\end{figure*}

As shown in Fig.~\ref{fig:fig_stretched_T1} and \ref{fig:fig_stretched_T2}, the slope for the $T_1$ decay is consistently 1/2, while the slope for the $T_2$ decay is consistently 3/2.
This holds true for NV centers both under and away from the BSCCO sample across the entire temperature range investigated.
We do note a minor deviation from this behavior at late times when the system is near a critical temperature (e.g., at 93~K for $T_1$ and 40~K for $T_2$).
The same phenomenon has been observed in a recent work~\cite{ziffer2024quantum}.

\suppsubsection{Validation of decay rate extraction methods}
The methodology used to extract the BSCCO-induced decay and decoherence rates, denoted as $\Gamma_1$ and $\Gamma_2$ in the main text, respectively, is outlined in Sections~I.2 and I.3.
In this subsection, drawing inspiration from a previous study~\cite{ziffer2024quantum}, we introduce an alternative method to validate the reliability of our findings.

\begin{figure*}[h!]
    \centering
    \includegraphics[width=0.7\textwidth]{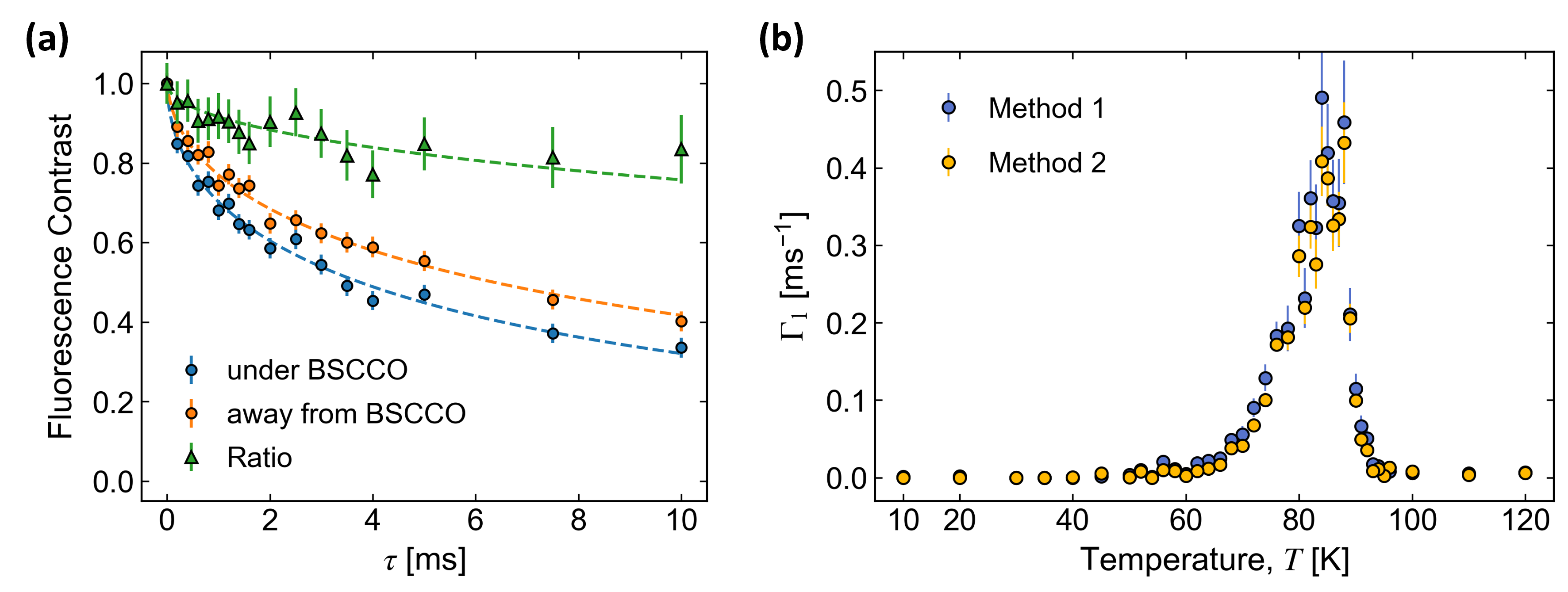}
    \caption{{\bf Comparing different methods of extracting the decay rate.}
    (a) The relaxation decays of NV centers under (blue) and away (orange) from BSSCO at $T=100$~K and $H=40$~G.
    The ratio (green) is calculated by dividing the blue data by the orange data.
    Error bars represent the photon shot noise, and the propagated $1\sigma$ uncertainty.
    All data are fitted with a stretched exponential decay with an exponent of 1/2.
    (b) The temperature dependence of the BSCCO-induced relaxation rate $\Gamma_1$ extracted using two methods.
    Error bars represent the propagated $1\sigma$ uncertainty.
    }
    \label{fig:fig_methods}
\end{figure*}

For each temperature, we measure the fluorescence contrast as a function of the time interval $\tau$, $C(\tau)$, for NV centers both under and away from BSCCO, and normalize the values to 1.
Taking the fluorescence measured at NV centers away from BSCCO as a reference, $C_\mathrm{ref}(\tau)$, we define a normalized decay:
\begin{equation}
    C_\mathrm{norm}(\tau) = \frac{C(\tau)}{C_\mathrm{ref}(\tau)}
    = e^{-\sqrt{\Gamma^*_1 \tau}}   ,
\end{equation}
where $\Gamma^*_1$ is the noise-induced decay rate.
An example of this is shown in Fig.~\ref{fig:fig_methods}(a).
By fitting $C_\mathrm{norm}(\tau)$ to a stretched exponential decay with an exponent of 1/2, we extract the temperature dependence of the BSCCO-induced decay rate $\Gamma^*_1(T)$, which is summarized as Method~2 in Fig.~\ref{fig:fig_methods}(b).
Compared to $\Gamma_1(T)$ extracted using the Method~1 in Section~I.2, $\Gamma^*_1(T)$ has the same asymmetry profile, with nearly identical peak amplitude and position.
This consistency underscores the robustness of our results.

\suppsubsection{Critical exponent analysis at zero field}
In this section, we provide a quantitative analysis of the critical exponent observed near the phase transition at zero magnetic field.
In particular, we perform a weighted least-squares fit on the data within the critical regime ($|T-T_c| \lesssim 6$~K) using the power law,
\begin{equation}
    \Gamma_1 = A~|T-T_c|^{-x},
\end{equation}
to extract the exponent $x$ for both superconducting (SC) and metallic (M) sides.
By fixing $T_c = 92.75$~K, we obtain $x_\mathrm{sc} = 1.10 \pm 0.09$ and $x_\mathrm{m} = 0.97 \pm 0.10$.
These values are in very close agreement with our claim that $x \approx 1$ on both sides of the transition, and $x \ne 0.5$ ($x \ne 0$) on the superconducting (metallic) side.
Crucially, this observation is robust when taking into account the uncertainty in the critical temperature.
By varying the assumed $T_c$ by $\pm 0.1$~K (determined by the $1\sigma$ uncertainty from our empirical Gaussian fit), we find the extracted exponent remains within the range $x_\mathrm{sc} \in [0.93, 1.25]$ and $x_\mathrm{m} \in [0.82, 1.10]$.
Moreover, if one treats $T_c$ as a free parameter and fits the dataset on both sides with an asymmetric power law, i.e., different $x$ for each side, it yields $T_c = 92.81 \pm 0.04$~K, $x_\mathrm{sc} = 1.19 \pm 0.12$, and $x_\mathrm{m} = 0.89 \pm 0.10$, still consistent with our findings.

\begin{figure*}[h!]
    \centering
    \includegraphics[width=0.36\textwidth]{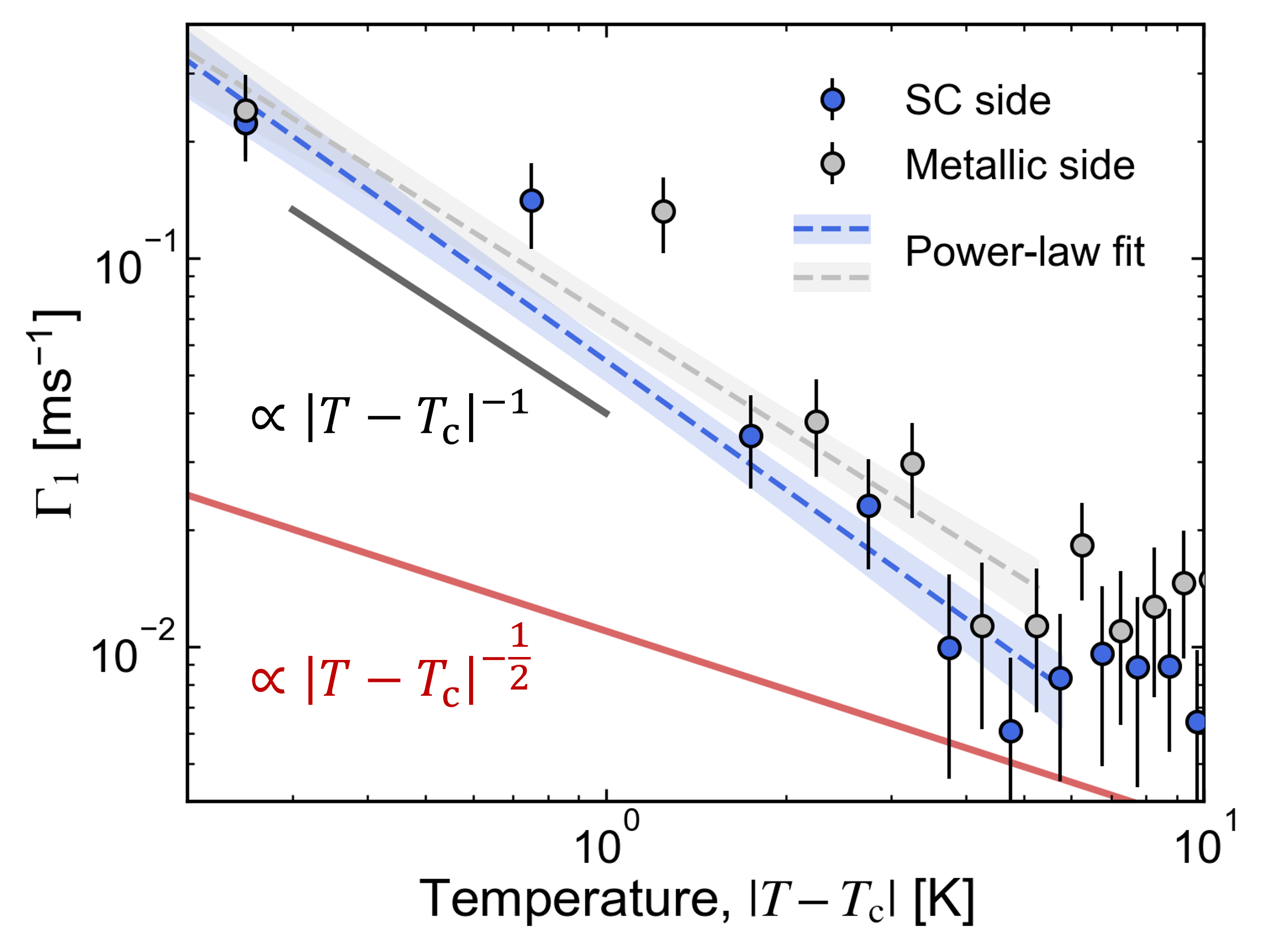}
    \caption{{\bf Critical exponent analysis.}
    $\Gamma_1(T)$ as a function of the absolute temperature deviation from the critical temperature, $|T - T_c|$, for both the metallic and superconducting (SC) sides of the transition, corresponding to Fig.~3(e) in the main text.
    A weighted least-squares fit is performed on the data, with dashed lines and shadow regions marking the best fits and $1\sigma$ confidence bands, respectively.
    Solid lines are guides of the eye showing distinct power-law scaling behaviors.
    }
    \label{fig:fig_critical_fit}
\end{figure*}

\begin{figure*}[h!]
    \centering
    \includegraphics[width=0.36\textwidth]{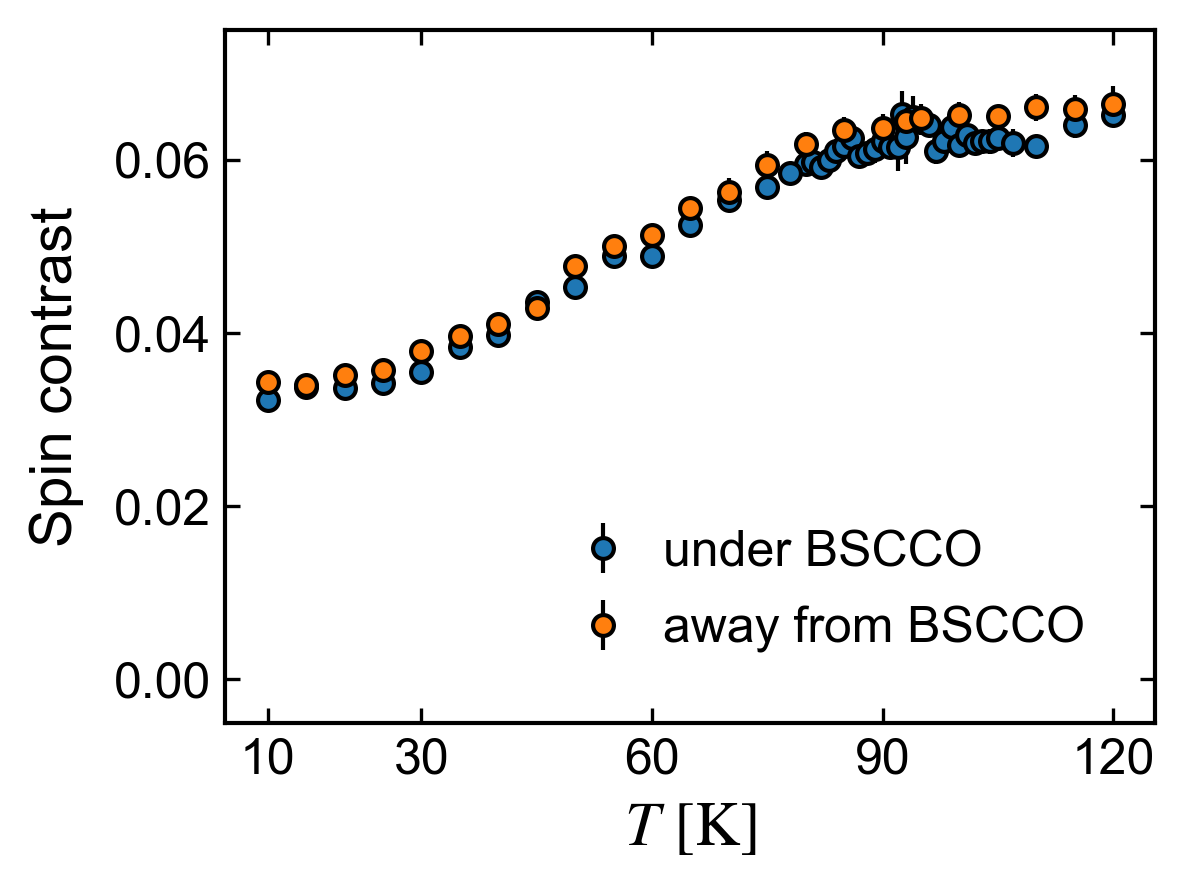}
    \caption{
    {\bf Spin contrast of relaxation decay as a function of temperature.}
    For both NV ensembles located under the BSCCO sample and far away from the sample.
    Error bars represent the $1\sigma$ uncertainty derived from the stretched exponential fits.
    }
    \label{fig:fig_contrast}
\end{figure*}

\suppsubsection{Temperature dependence of spin contrast}
During the experiments, we have observed that the spin contrast of NV centers decreases at low temperatures.
Specifically, the contrast begins to drop below 90~K and stabilizes around 30~K, see Fig.~\ref{fig:fig_contrast}.
Such a decrease has been observed in multiple diamond samples (not used in this work) and reported in other independent works~\cite{ernst2023temperature, happacher2023temperature}.

As this decrease in spin contrast will effectively lower the overall signal-to-noise ratio (SNR), we intentionally increase the measurement averaging time at lower temperatures in order to ensure a consistent data quality across the entire temperature range.
The longer acquisition compensates for the lower signal contrast, thereby maintaining a relatively consistent SNR.
In addition, even at its minimum, the remaining spin contrast of over 3\% is still substantial enough for robust and reliable extraction of the decay timescales.

\suppsubsection{Langevin dynamics modeling of finite-field noise}
As discussed in the Methods, the depolarization rate is dominated by magnetic field fluctuations. We may relate the fluctuations to two-point displacement correlation functions of pairs of vortices. Under the assumption of uncorrelated pairs and identical statistical behavior, the $T_1$ noise is effectively proportional to the two-point displacement correlators $\Gamma_1 =3 (g \mu_B/\hbar)^2 \langle\langle\mathcal{N}_{xx}\rangle\rangle \propto \langle \u_{\Omega} \cdot \u_{-\Omega}\rangle_T$, with a proportionality constant given by the product of a geometric factor and the average vortex density.
We model the vortex motion at GHz frequency by Langevin dynamics \cite{blatter1994vortices,cheng2026qubit}, for a characteristic vortex displacement $\u_t\equiv\u(t)$,
\begin{equation}
    \label{eq:vortex_langevin}
    \mu_v \ddot{\u}_t=n_s h\left(\dot{\u}_t \times \hat{\z}\right)-K \u_t-\eta
    \dot{\u}_t+\bm{\zeta}_t \,.
\end{equation}
where $\mu_v$ stands for vortex mass per unit length; the superfluid density $n_s=m^*(e^{* 2} \mu_0 \lambda^2)^{-1}$, and the Magnus force is $n_s h\left(\dot{\u}_t \times \hat{\z}\right)$; $K$ denotes the effective restoring spring constant per unit length, which originates primarily from vortex pinning (the Labusch parameter \cite{labusch1968elasticity}); a phenomenological drag coefficient $\eta$ arises mainly due to viscosity; the stochastic force $\bm{\zeta}$ satisfies the fluctuation-dissipation theorem,
\begin{equation}
    \langle\zeta_t^\alpha \zeta_{t^{\prime}}^\beta\rangle_T=2 (\eta/a_c) k_B T \delta^{\alpha \beta}
    \delta\left(t-t^{\prime}\right),
\end{equation}
where $a_c=1.55~\mathrm{nm}$ represents the interlayer spacing between adjacent active conducting $\mathrm{CuO}_2$ bilayers.

To find the two-point displacement correlation functions, we Fourier transform Eq.~\eqref{eq:vortex_langevin} using $\u_t=\u_{\Omega} e^{-i\Omega t}$. The corresponding Langevin equation in frequency space becomes:
\begin{equation}
    \left(
    \begin{array}{cc}
        -\mu_v\Omega^2-i\eta\Omega+K & i n_s h\Omega \\
        -i n_s h\Omega & -\mu_v\Omega^2-i\eta\Omega+K
    \end{array}
    \right)\u_{\Omega}=\bm{\zeta}_{\Omega} \,.
\end{equation}
where the $\mu_v \Omega^2$ term can be neglected, since the experimentally relevant frequency regime corresponds to a massless limit, $\mu_v\Omega \ll n_s h, \eta$. Based on Eqs. (M4), (M19), the noise can be written as
\begin{equation}
    \begin{aligned}
        \Gamma_1
        &= \frac{3 n_v}{2}\left(\frac{g\mu_B}{\hbar}\right)^2
        \langle \u_{\Omega}\cdot\u_{-\Omega}\rangle_T
        \int_{\text{sample}}\left|\bm{g}(\bm{R})\right|^2, \\
        \langle \u_{\Omega}\cdot\u_{-\Omega}\rangle_T
        &= \frac{
            \left(2\eta/a_c\right) k_B T
            \left\{K^2+\left[\eta^2+\left(n_s h\right)^2\right]\Omega^2\right\}
        }{
            \left\{K^2-\left[\eta^2+\left(n_s h\right)^2\right]\Omega^2\right\}^2
            + 4\eta^2\Omega^2 K^2
        }
        = \frac{2\bar{\eta} k_B T}{a_c n_s h\Omega^2}
        \frac{\bar{K}^2+1+\bar{\eta}^2}{
            \left(\bar{K}^2-1-\bar{\eta}^2\right)^2
            + 4\bar{K}^2\bar{\eta}^2
        } \,,
    \end{aligned}
    \label{eq:Gamma1_TH}
\end{equation}
where we have defined the dimensionless quantities: $\bar{K}=K/(n_s h \Omega)$ and $\bar{\eta}=\eta/(n_s h)$. In the $\bar{K},\bar{\eta}\gg1$ limit, $\Gamma_1\sim \bar{\eta}(\bar{K}^2+\bar{\eta}^2)^{-1}$; in the $\bar{K}\gg1,\bar{\eta}$ limit, $\Gamma_1\sim\bar{\eta}\bar{K}^{-2}$.

In practice, the geometric factor $\int\left|\bm{g}(\bm{R})\right|^2$ is determined using the single-vortex magnetic field profile $\B_0$ (Carneiro--Brandt model \cite{carneiro2000vortex}) for a sample with thickness $d\approx200~\mathrm{nm}$. Here, the gradient $\g$ is defined in Eq.~(M17) in the Methods as $\g(\R)=-\left.\bm\nabla_{\parallel}\left[B_{0}^\rho(\bm\rho, z_0)(\hat{\bm{\rho}} \cdot \hat{\bm{x}})\right]\right|_{\bm{\rho}=\bm{\rho}_0-\bm{R}}=-\cos \theta \partial_\rho B_0^\rho \hat{\bm{\rho}}+\sin \theta \rho^{-1} B_0^\rho \hat{\bm{\theta}}$, $\cos\theta=\hat{\bm{\rho}} \cdot \hat{\bm{x}}$. The explicit expressions for the field profile and the corresponding geometric factor integrals are given by:
\begin{equation}
    \begin{aligned}
        B_0^\rho\left(\rho,z_0\right)
        &= \frac{\Phi_0}{2\pi\lambda^2}
        \int_0^{\infty} d k\,
        \frac{k J_1(k\rho)c_1(k)e^{-k z_0}}{k^2+\lambda^{-2}}, \quad
        c_1(k)
        = \frac{q\tanh(q d/2)}{q\tanh(q d/2)+k},
        \quad q^2\equiv k^2+\lambda^{-2}, \\
        \int_{\text{sample}}\left|\bm{g}(\bm{R})\right|^2
        &= \int_{\mathbb{R}^2}\left|\bm{g}(\bm{R})\right|^2
        - \int_{\mathbb{R}^2\backslash\text{sample}}\left|\bm{g}(\bm{R})\right|^2,
        \text{ where } \int_{\mathbb{R}^2}\left|\bm{g}(\bm{R})\right|^2
        = \frac{\Phi_0^2}{4\pi\lambda^4}
        \int_0^{\infty} d k\,
        \frac{k^3 c_1(k)^2 e^{-2 k z_0}}{\left(k^2+\lambda^{-2}\right)^2}.
    \end{aligned}
    \label{eq:geometric_factor}
\end{equation}
Although this evaluation formally depends on the sample geometry along with the NV center's lateral and vertical ($z_0=40~\mathrm{nm}$) positions relative to the boundary as depicted in Fig.~3(a), its scaling is dominated by the NV depth and the temperature, since the sample integral closely approximates the infinite-plane integral.

Now, let us extract the effective restoring spring constant $K(T)$ from the finite-field depolarization rate $\Gamma_1(T, H)$. The reference quantity $n_s h\Omega$ can be calculated using the penetration depth $\lambda(T)$ in Ref.~\cite{jacobs1995penetration} and an NV angular frequency $\Omega \approx 2\pi \times D_\mathrm{gs} = 18.03~\mathrm{Grad/s}$. 
The vortex viscosity can be estimated from the relation $\eta = \Phi_0 H_{c2}(T)/\rho_n$ according to the Bardeen--Stephen model~\cite{bardeen1965motion}, where the upper critical field is evaluated with the empirical formula $H_{c 2}(T)=\Phi_0 /\left(2 \pi \xi_0^2\right) [1-\left(T / T_c\right)^2]$, $\xi_0 = 1.6~\text{nm}$; the normal-state resistivity is extrapolated from measurements in Ref.~\cite{aichner2025enhanced}. 
Fitting $\Gamma_1(T, H)$ using Eqs.~\eqref{eq:Gamma1_TH},\eqref{eq:geometric_factor} with the estimated $\bar{\eta}(T)$ provides a reasonable estimate of $K(T)$.

\begin{figure*}[h!]
    \centering
    \includegraphics[width=0.7\textwidth]{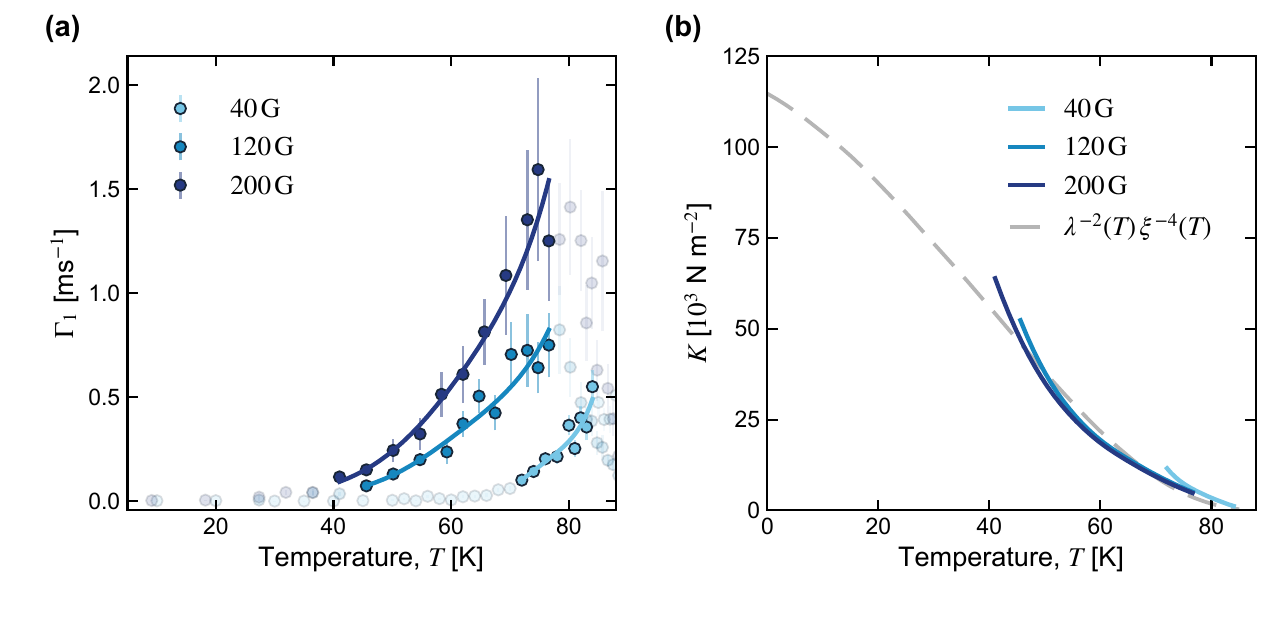}
    \caption{{\bf Extraction of the effective spring constant $K(T)$.}
    (a) Measured $\Gamma_1(T)$, at applied field $H=40, 120, 200~\mathrm{G}$.
    Solid curves show fits of the increasing part of the noise to the vortex Langevin dynamics model.
    (b) $K(T)$ extracted from the fits in (a); the dashed line indicates the scaling $\lambda^{-2}(T) \xi^{-4}(T)$.
    Note that $K$ represents the averaged effective restoring spring constant within the dynamical model rather than a direct measurement of the microscopic pinning scale.
    }
    \label{fig:fig_Gamma1_fit_K}
\end{figure*}

The extracted values of $K(T)$ are comparable to those obtained for optimally doped La$_{2-x}$Ce$_{x}$CuO$_4$~\cite{zuev2003vortex} and YBa$_2$Cu$_3$O$_{7-\delta}$~\cite{pesetski2000experimental}, which were also extracted in the strongly-pinned Campbell limit, $\bar{K}\gg\bar{\eta}$ \cite{srpcic2020flux}. Optimally doped cuprates typically exhibit a larger pinning than non-optimally doped variations due to their enhanced superfluid density or equivalently larger condensation energy scale. The restoration effect is stronger at lower temperatures or low fields [Fig.~\ref{fig:fig_Gamma1_fit_K}], where pinning dominates the vortex-vortex interactions. The weaker restoring effect at higher temperatures indicates the potential emergence of a depinned vortex solid or a vortex liquid \cite{beidenkopf2005equilibrium}.

An intuitive understanding of the pinning can be gained from a rotationally symmetric ansatz \cite{pesetski2000experimental} for an effective pinning potential per unit length $U(R)$, satisfying the conditions that its curvature equals the restoring constant at the bottom of the potential well $\left.\partial_R^2 U \right|_{R=0}=K_0$, and vanishes at the inflection point $\left.\partial_R^2 U \right|_{R=\xi}=0$. Expanding the potential in even powers of radius $R$ yields
\begin{equation}
    U(T,R)
    = \frac{1}{2}K_0(T)\xi^2(T)\tilde{R}^2
    - \frac{1}{12}K_0(T)\xi^2(T)\tilde{R}^4
    + \mathcal{O}(\tilde{R}^6),
    \text{ where }\tilde{R}
    = R/\xi(T).
\end{equation}
The pinning energy per unit length $E_p$ cannot be determined from this local potential alone. Following Ref.~\cite{pesetski2000experimental}, we adopt the phenomenological estimate $E_p(T)\simeq2U(T,\xi)\simeq\frac{5}{6}K_0(T)\xi^2(T)$, which relates the restoring constant to the pinning-energy scale.
For a cylindrical defect, neglecting kinetic-energy savings, the pinning energy per unit length is estimated from the condensation energy saved within its circular cross-section of effective radius $r_d$:
\begin{equation}
    E_p(T)\simeq\left(\frac{H_c^2(T)}{2\mu_0}\right)(\pi r_d^2)
    \implies K_0(T)\simeq
    \frac{3\Phi_0^2 r_d^2}{40\pi\mu_0\lambda^2(T)\xi^4(T)},
\end{equation}
where $H_c=\Phi_0/(2\sqrt{2}\pi\lambda\xi)$ is the thermodynamic critical field.
Under these assumptions, fitting the extracted $K(T)$ curves to the scaling $\lambda^{-2}(T)\xi^{-4}(T)$, as shown in Fig.~\ref{fig:fig_Gamma1_fit_K}, yields $r_d\approx0.79~\mathrm{nm}$, approximately half the assumed zero-temperature coherence length $\xi_0=1.6~\mathrm{nm}$. If the kinetic energy saved in the surrounding superfluid also contributes to the pinning energy, the estimated $r_d$ can be smaller. The subnanometer defect may originate from local disorder.

\suppsection{Supplementary Results}

\suppsubsection{Critical scaling of noise in a second BSCCO flake}
In the main text, we observed critical fluctuations at the superconducting phase transition via depolarization ($T_1$) spectroscopy.
To confirm that our observation of this critical behavior is robust and consistent, we perform the same measurements on a second, freshly exfoliated BSCCO flake with a slightly smaller thickness of $\sim$150~nm.

\begin{figure*}[h!]
    \centering
    \includegraphics[width=1\textwidth]{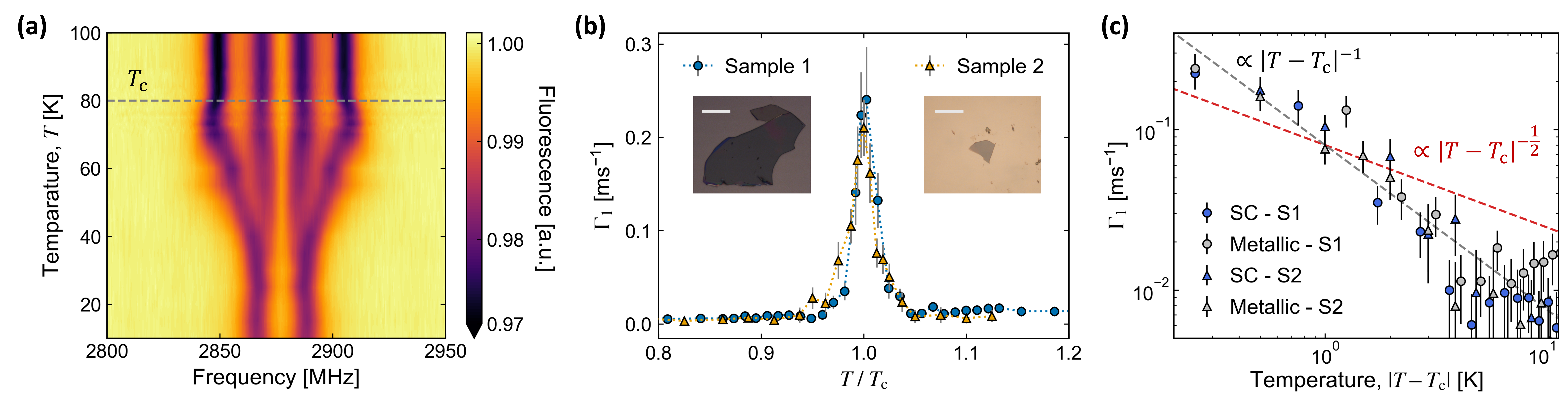}
    \caption{{\bf Repeated experiments on a freshly exfoliated BSCCO sample.}
    (a) The ODMR spectrum of NV centers under a second BSCCO flake with an applied field of $H=10$~G at temperatures from 10~K to 100~K.
    The gray dashed line marks the transition temperature $T_c \approx 80$~K of the flake.
    (b) BSCCO-induced decay rate of NVs, $\Gamma_1(T)$, for the original sample used in the main text (sample 1, S1) and a second BSCCO sample (sample 2, S2), without any external magnetic field.
    Inset: optical images of the two samples with scale bars of 20~$\mu$m.
    Variations in apparent color result from different imaging conditions.
    (c) Critical analysis of $\Gamma_1$ near transition for the fresh sample.
    For both the metallic (m) and superconducting (sc) sides of the transition, the data are consistent with a power-law behavior of $\Gamma_1\propto|T-T_c|^{-x}$ with $x=1$, and clearly deviate from BCS mean-field prediction ($x_{\rm sc}=1/2$ and $x_{\rm m} = 0$).
    }
    \label{fig:fig_sample2}
\end{figure*}

First, by measuring the ODMR spectrum from 10~K to 100~K under an external magnetic field of 10~G, we identify a transition temperature of $T_c \approx 80$~K for this flake [Fig.~\ref{fig:fig_sample2}(a)].
Compared to the primary sample used in the main text, this $T_c$ is slightly lower --- such a variation is expected as $T_c$ is quite sensitive to the precise dopant density in high-$T_c$ cuprates.
However, although the exact value of $T_c$ may be different across samples, the universal features associated with the superconducting transitions, specifically the critical exponents, are expected to be robust.

To investigate the critical fluctuations near the transition, we perform $T_1$ spectroscopy with a denser temperature sampling, and extract the BSCCO-induced decay rate $\Gamma_1$ following the same procedure described previously.
As shown in Fig.~\ref{fig:fig_sample2}(b), both the peak amplitude and peak width of the data measured on the second flake (sample 2) are comparable with those of the original flake (sample 1), when scaled by $T_c$.
At the same time, the peak positions are consistent with transition temperatures for each sample, showing the consistency in our observations.
More importantly, plotting $\Gamma_1$ on a log-log scale in Fig.~\ref{fig:fig_sample2}(c) reveals that data at both the superconducting and metallic sides of the transition clearly deviate from the BCS mean-field prediction ($x=1/2$ and $x=0$), and are consistent with our theoretical prediction of $x=1$ from critical fluctuations.
Applying the same power-law analysis, the critical exponents are extracted as $x = 1.05 \pm 0.15$ on the superconducting side and $x = 1.11 \pm 0.16$ on the metallic side, while remaining within the range of $[0.95, 1.14]$ and $[1.01, 1.21]$ respectively by varying the assumed $T_c$ by $\pm 0.1$~K.

To complete our analysis, we again extract the scattering timescales for both metallic and superconducting sides in sample 2, following an identical procedure as for sample 1 but adjusting for the number of layers because of reduced thickness (the details are discussed in the Methods section (2.3.1) in the main text).
Unlike critical exponents, these scattering timescales are not universal, and are expected to vary between different samples. 
Indeed, sample 2 exhibits $\tau_{\rm sc}/ \tau_{\rm GL} = 0.65 \pm 0.08$, $\tau_{\rm m}/ \tau_{\rm GL} = 2.82 \pm 0.30$, which are comparable to, but not exactly the same as, the values we extracted for sample 1: $\tau_{\rm sc}/ \tau_{\rm GL} = 0.32 \pm 0.03$, $\tau_{\rm m}/ \tau_{\rm GL} = 2.11 \pm 0.22$.
Nevertheless, our main conclusion --- that the critical scattering timescales in thin-film BSCCO are comparable to the weak coupling prediction $\tau^{\rm wc}_{\rm sc[m]} / \tau_{\rm GL} = 0.5[1]$ --- is robust across both samples. 
These consistent observations across different samples confirm that such critical fluctuations are a robust, intrinsic feature of the superconducting transition in BSCCO.

\begin{figure*}[h!]
    \centering
    \includegraphics[width=0.7\textwidth]{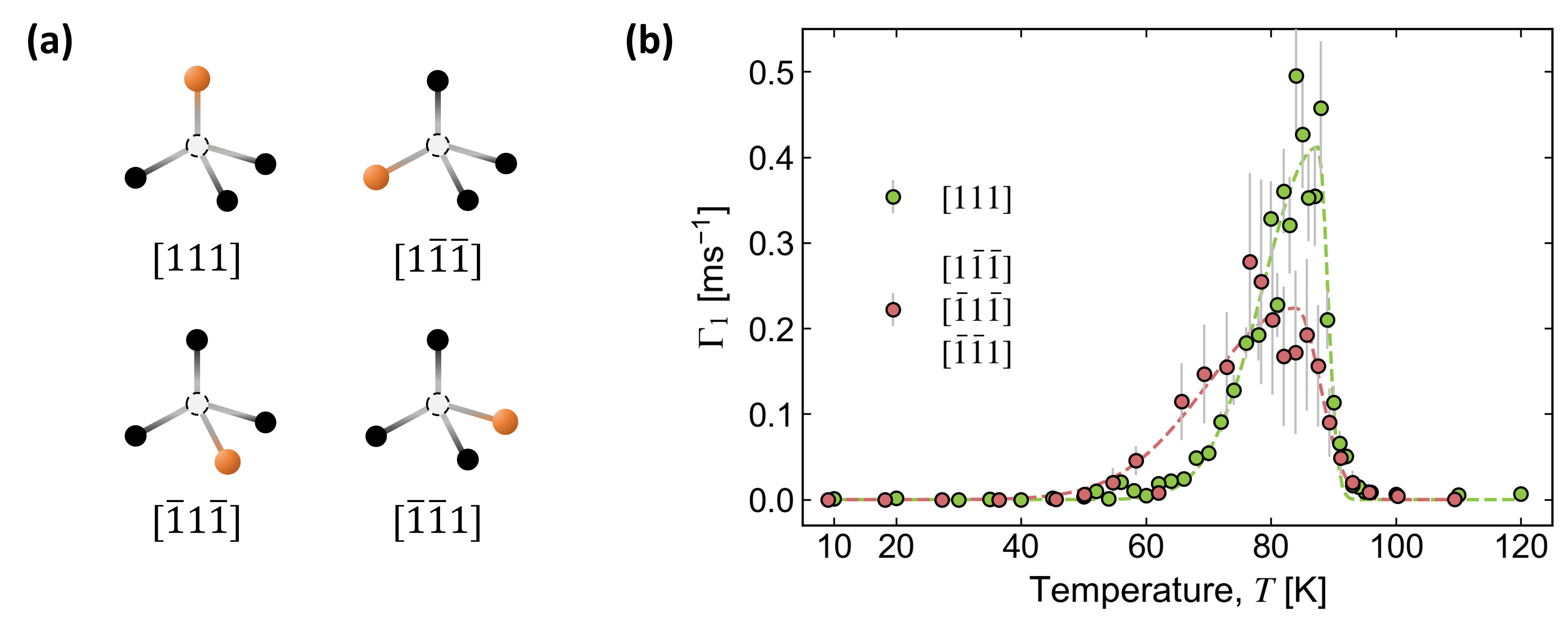}
    \caption{{\bf Off-axis $T_1$ relaxometry.}
    (a) Schematic of all four NV orientations in a single-crystal diamond lattice, with the [111] group aligned along the out-of-plane direction (surface normal).
    (b) BSCCO-induced relaxation rate, $\Gamma_1$, as a function of temperature for the [111] NV group (green) and the three degenerate non-[111] NV groups (red), measured under a bias field $H=40$~G.
    Error bars represent the propagated $1\sigma$ uncertainty, while the curves correspond to the asymmetric Gaussian fits.
    The consistency in peak position implies the isotropic nature of the critical current fluctuations.
    }
    \label{fig:fig_offaxis}
\end{figure*}

\suppsubsection{Probing critical fluctuations in multiple directions}
Relaxation-based ($T_1$) noise spectroscopy of NV centers is primarily sensitive to the components of the fluctuating magnetic field perpendicular to the NV quantization axis.
For our primary NV axis (aligned along the $[111]$ direction, normal to the sample surface), this corresponds to sensitivity to in-plane magnetic noise, $B_x(t)$ and $B_y(t)$.
To investigate the full magnetic noise tensor --- specifically, components corresponding to out-of-plane fluctuations $B_z(t)$ --- we perform relaxation measurements on the off-axis NV centers.
These correspond to the three equivalent crystal orientations ($[1\bar{1}\bar{1}]$, $[\bar{1}1\bar{1}]$, $[\bar{1}\bar{1}1]$), which are canted with respect to the surface normal, as illustrated in Fig.~\ref{fig:fig_offaxis}(a).
In our setup geometry, the [111] NV group is aligned along the out-of-plane ($z$) direction, while the other three groups form an angle $\theta=\arcsin(1/3) \approx 19.5^\circ$ with the sample plane.
Consequently, the [111] group serves as a sensor for in-plane magnetic field fluctuations, whereas the non-[111] groups exhibit a significant projection perpendicular to the $z$-axis, allowing them to detect out-of-plane components as well.

In the absence of an external magnetic field, all four NV orientations are degenerate, and the sensor is equally responsive to signals in all directions.
When a bias field is applied along the out-of-plane direction, the [111] group exhibits a larger Zeeman splitting due to its maximal field projection, while the off-axis groups remain degenerate with a smaller splitting.
By selectively driving the appropriate resonance frequencies, we can independently address and measure the relaxation dynamics of either the on-axis [111] or the off-axis non-[111] NV populations.

Figure~\ref{fig:fig_offaxis}(b) shows the temperature dependence of the BSCCO-induced relaxation rate $\Gamma_1$ for both the $[111]$ and non-$[111]$ groups under a bias field of $H=40$~G.
We observe that the peak positions of $\Gamma_1(T)$ are consistent for both the on-axis NVs and the off-axis NVs.
This consistency implies that the critical fluctuations generate time-varying magnetic fields with significant components in all spatial directions.

\begin{figure*}[h!]
    \centering
    \includegraphics[width=0.6\textwidth]{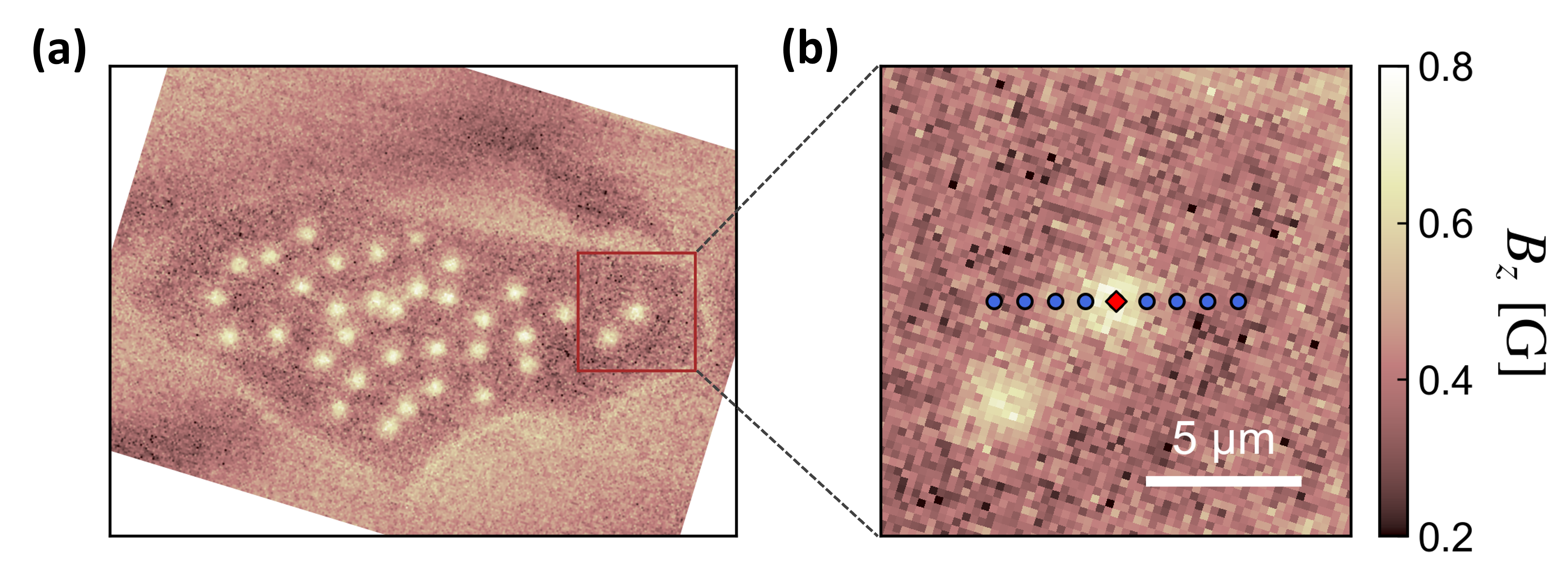}
    \caption{{\bf Imaging the isolated pinned vortices.}
    (a) The local magnetic field $B_z$ map measured at $T = 40$~K under a small magnetic field of $H \approx 0.5$~G.
    (b) A zoom-in image of the square area in (a), same as Fig.~4(d) left panel in the main text.
    The dots mark the positions across an isolated vortex, used in the line-cut measurements.
    }
    \label{fig:fig_vortex_iso}
\end{figure*}

\suppsubsection{Imaging of an isolated pinned vortex}
In this section, we provide additional details of the isolated vortex discussed in Fig.~4(d) in the main text.
We perform widefield ODMR measurements on a region containing the BSSCO flake to image the local magnetic field.
To spatially resolve individual vortices, we apply a very low magnetic field ($H \approx 0.5$~G) and field-cool the system to $T = 40$~K, well below the transition temperature.
As shown in Fig.~\ref{fig:fig_vortex_iso}, under these conditions, the vortices are spatially stable and well-separated, allowing for further spatially resolved investigation of noise, such as measurements along line-cuts across vortices.

\begin{figure*}[h!]
    \centering
    \includegraphics[width=0.55\textwidth]{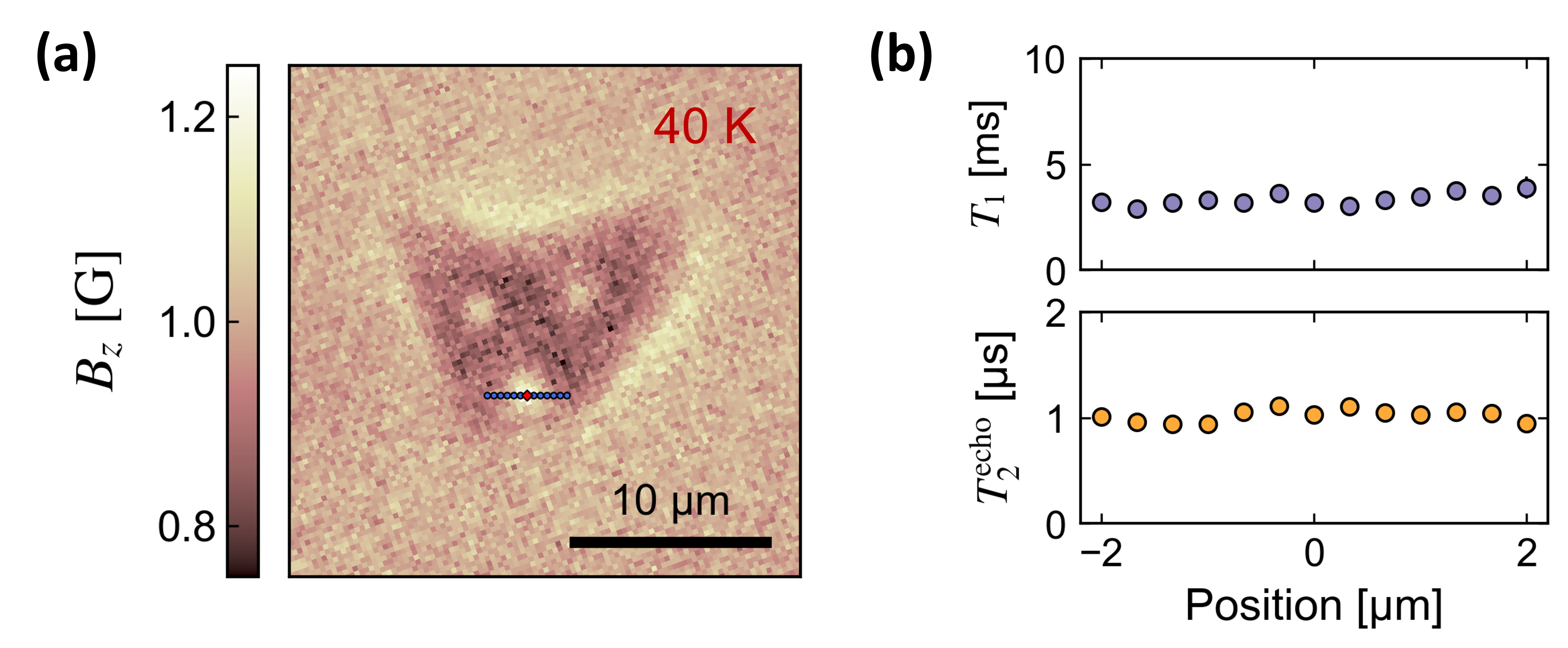}
    \caption{{\bf Noise measurements across an isolated pinned vortex in Sample 2.}
    (a) Isolated vortices imaged in BSSCO Sample 2 under $H \approx 1~$G at 40~K.
    The dots mark the positions measured across the pinned vortex near the edge of the flake.
    (b) Measured relaxation time $T_1$ and decoherence time $T_2^\mathrm{echo}$ of NV centers at different spatial positions with the vortex core centered at zero.
    Error bars represent the 1$\sigma$ uncertainty from fitting the corresponding $T_1$ and $T_2$ decay curves.
    }
    \label{fig:fig_linecut}
\end{figure*}

\suppsubsection{Magnetic noise from the vortex core}
To better support the experimental results shown in Fig.~4(d) in the main text, we repeat the $T_1$ and $T_2$ noise measurements across an isolated pinned vortex in the second BSCCO flake (Sample 2) with a finer spatial sampling of 1/3~$\mu$m.
As shown in Fig.~\ref{fig:fig_linecut}, neither $T_1$ nor $T_2^\mathrm{echo}$ exhibits a systematic appreciable variation as the measurement position crosses the vortex core.
These measurements are therefore consistent with our previous results in Fig.~4(d).
We note that, although the spatial sampling interval can be chosen to be smaller, the effective spatial resolution of our confocal setup is limited by the optical diffraction limit, which is approximately 1~$\mu$m.
Consequently, signals with spatial spacing below this limit will inevitably be averaged with the surrounding region.

This absence of the change in $T_1$ and $T_2$ observed on a line-cut across a pinned vortex indicates that the amplitude of the noise from the vortex core is too weak to be measured with our probe.

\suppsubsection{Magnetic field dependence of vortex density}
In Fig.~2(d) of the main text, we have presented a static vortex map acquired under a bias field of $H\approx2$~G.
To validate this observation, we perform a quantitative analysis comparing the measured vortex density against theoretical expectations.

In type-II superconductors, magnetic flux is quantized, with each individual vortex carrying a single flux quantum $\Phi_0 = h/(2e) \approx 20.68~{\rm G \cdot \mu m^2}$.
Ideally, the average vortex density (flux density) within the sample is equal to the externally applied magnetic field $H$.
The expected number of superconducting vortices, consequently, is given by:
\begin{equation}
    N = \frac{H \cdot S}{\Phi_0}    ,
\end{equation}
where $S$ is the effective sample area containing the vortices.

\begin{figure*}[h!]
    \centering
    \includegraphics[width=0.9\textwidth]{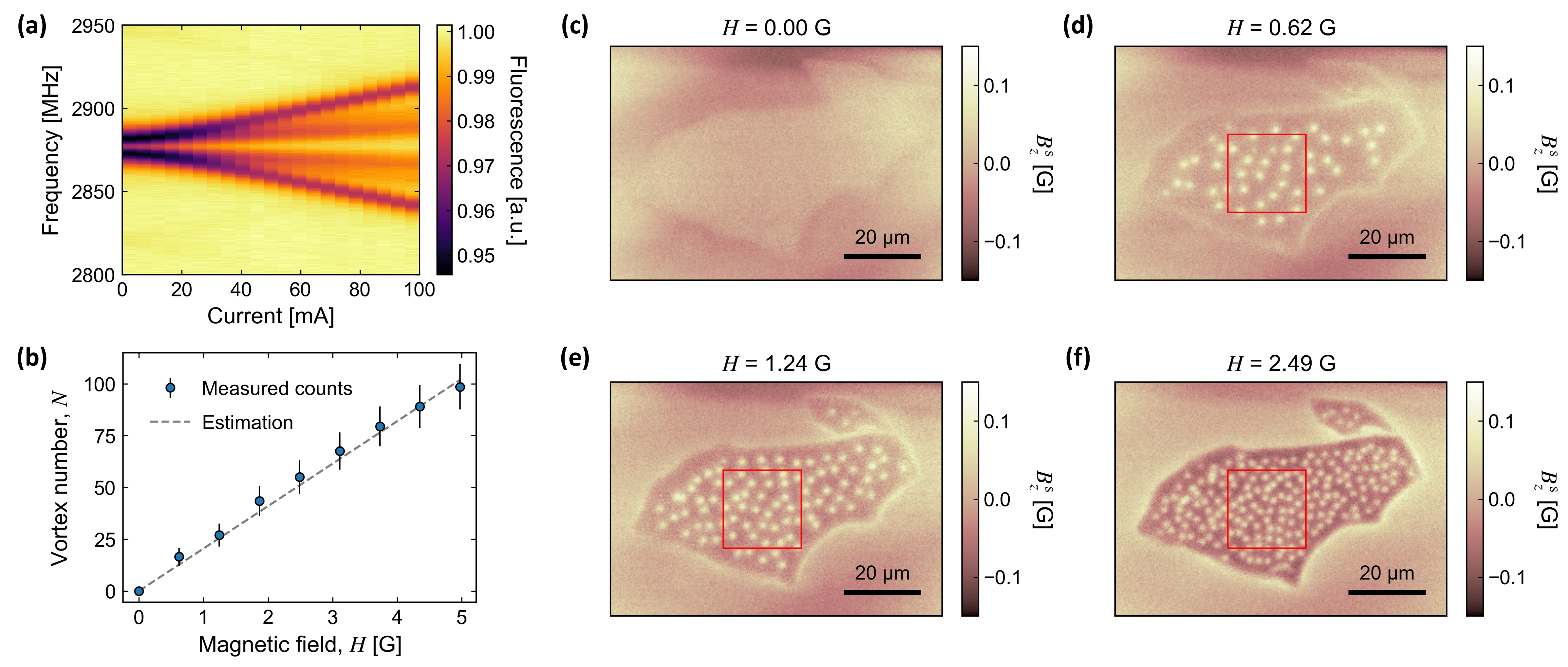}
    \caption{{\bf Dependence of vortex number on the applied magnetic field.}
    (a) ODMR spectrum versus the applied current.
    Measurements are performed on NV centers located away from the BSCCO flake, at 120~K.
    (b) Vortex number $N$ as a function of the bias magnetic field $H$.
    The dashed line represents the theoretical estimation, while the points represent experimentally measured counts.
    Error bars represent one standard deviation of the boundary vortex counting.
    (c-f) Maps of the stray magnetic field $B_z^\mathrm{s}$ beneath the BSCCO sample at representative bias fields.
    Measurements are performed at 50~K, after a separate field-cooling process for each bias field.
    }
    \label{fig:fig_vortex_number}
\end{figure*}

The external magnetic field $H$ here is generated by a solenoid magnet, driven by a DC power supply.
We calibrate it by applying variable current and measuring the ODMR spectrum at NV centers located away from the BSCCO flake, as shown in Fig.~\ref{fig:fig_vortex_number}(a).
The effective sample area $S$ is defined as a square region on the flake [red squares in Fig.~\ref{fig:fig_vortex_number}(d-f)], determined via pixel counting from the optical image.
The resulting theoretical estimation for the field dependence of the vortex number is plotted as the dashed line in Fig.~\ref{fig:fig_vortex_number}(b).

In experiments, we perform widefield ODMR measurements at 50~K, after a separate field-cooling process for each distinct bias field.
We then extract the stray magnetic field $B_z^\mathrm{s}$; representative maps are shown in Fig.~\ref{fig:fig_vortex_number}(c-f).
To determine the vortex number, we count vortices located fully within the square region as 1, while those intersecting the boundaries are weighted as 1/2.
The resulting counts are summarized in Fig.~\ref{fig:fig_vortex_number}(b), with error bars corresponding to one standard deviation of the boundary vortex counting.
We find excellent quantitative agreement between the observed vortex counts and the theoretical estimation.
Crucially, the measured counts scale linearly with the applied magnetic field.
This observation confirms that the detected magnetic features are indeed individual vortices carrying single flux quanta, behaving as expected for a type-II superconductor.

\suppsubsection{Temperature evolution of the vortex phases}
In the main text, we have observed evidence of vortex liquid and solid phases using our noise spectroscopy.
Here, we provide additional evidence of these observations by directly visualizing the vortex evolution through static magnetic field imaging.

\begin{figure*}[h!]
    \centering
    \includegraphics[width=0.8\textwidth]{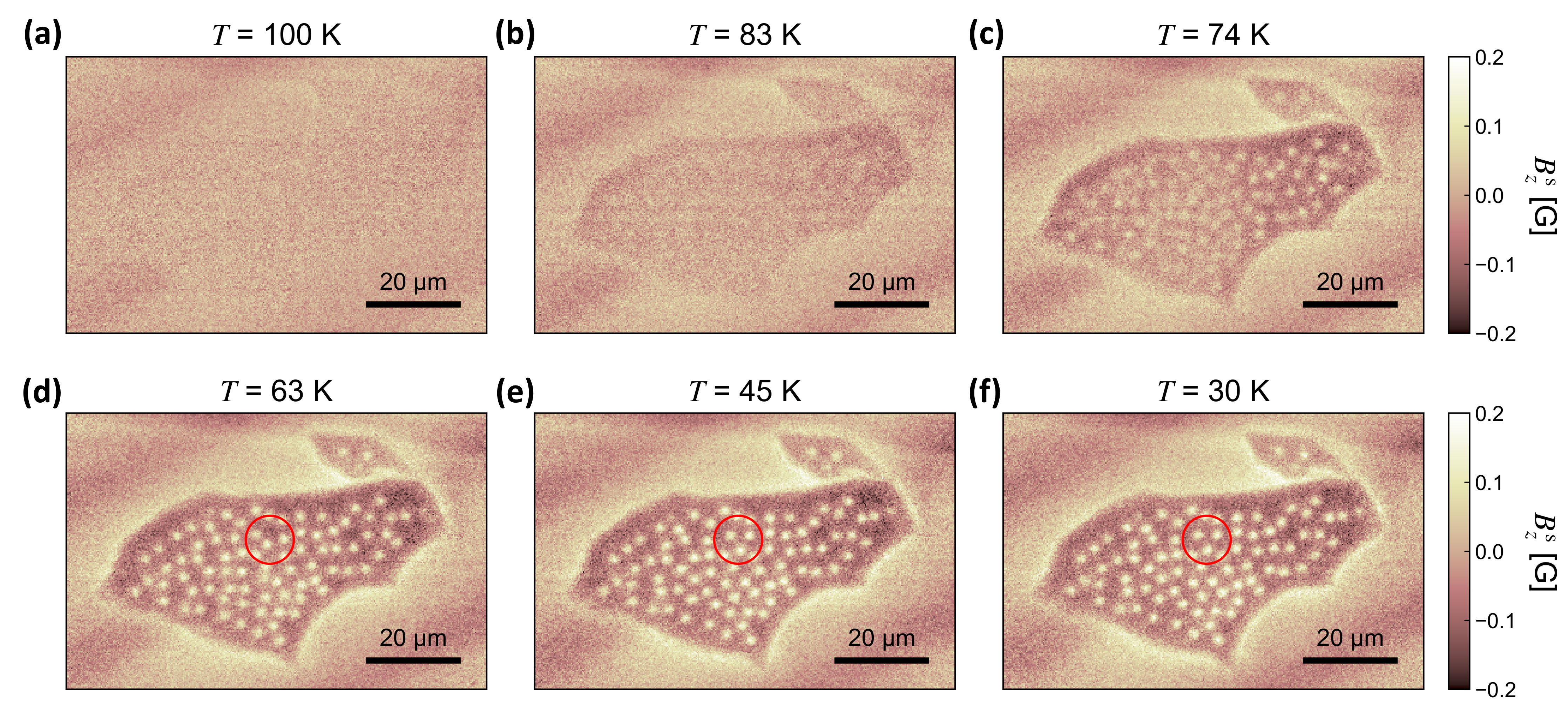}
    \caption{{\bf Vortex evolution with temperature.}
    Stray magnetic field ($B_z^\mathrm{s}$) maps extracted at representative temperatures during field-cooling, with an external field of $H \approx 1.24$~G.
    The images show the evolution from (a) the metallic phase, through (b, c) the vortex liquid phase, to (d-f) the vortex solid phase.
    While vortices are generally stable in the solid phase, some vortices could jump to different pinning sites between 75~K and 45~K, as highlighted by red circles in (d-f).
    }
    \label{fig:fig_vortex_temperature}
\end{figure*}

Similar to the static vortex map shown in Fig.~2(d) in the main text, we perform widefield ODMR measurements under a small bias magnetic field of $H \approx 1.24$~G.
During the field-cooling process, we extract the stray magnetic field $B_z^\mathrm{s}$ map at each temperature by fitting the full ODMR spectrum.
We note that each map takes an acquisition time of approximately 40~minutes for a high signal-to-noise ratio (SNR).
One can shorten this time to some extent by accepting a lower SNR.
As shown in Fig.~\ref{fig:fig_vortex_temperature}, the evolution of the vortices with temperature change can be summarized as follows:
\begin{itemize}
    \item \textbf{Metallic phase:} above 90~K, no magnetic features are discernible, consistent with the sample being in the normal metallic state.
    \item \textbf{Vortex liquid phase:} below $\sim 90$~K, the outline of the BSCCO flake becomes visible due to the Meissner effect, confirming the transition to the superconducting state.
    However, individual vortices are absent. In this liquid phase, vortices are highly mobile, and their rapid motion averages out to a uniform background over the acquisition time.
    \item \textbf{Vortex solid phase:} below $\sim 75$~K, individual vortices begin to resolve as the system enters the vortex solid phase.
    \begin{itemize}
    \item \textbf{Pinning dynamics:} crucially, in the intermediate range between 75~K and 45~K, we observe specific vortices ``hopping" between pinning sites (highlighted by red circles in Fig.~\ref{fig:fig_vortex_temperature}).
    \item \textbf{Frozen solid:} below $\sim 45$~K, the vortex structure becomes completely frozen, with no further changes in vortex positions observed.
    \end{itemize}
\end{itemize}

These static imaging results are consistent with the noise measurements presented in the main text.
We emphasize that, however, this static field imaging method is effective only in the low-field regime.
Since the vortex density increases linearly with the applied magnetic field, individual vortices become unresolvable at higher fields ($H \gtrsim 10$~G).
In contrast, our noise spectroscopy approach with $T_1$ and $T_2$ measurements detects the magnetic field fluctuations rather than the static profile.
Thus, it can distinguish between vortex phases in high-field regimes where static imaging is not applicable, making it uniquely suited to detecting the rapid dynamics characteristic of the vortex phases.

\bookmarksetupnext{level=subsection}
\putbib[references]

\end{bibunit}

\end{document}